\documentclass[journal]{IEEEtran}
\usepackage{cite}
\usepackage{amsmath}
\usepackage{algorithmic}
\usepackage{array}
\usepackage{url}
\usepackage{xcolor}
\usepackage{graphicx}
\usepackage{multirow}
\usepackage{amsmath}
\usepackage{amsfonts}
\usepackage{verbatim}
\usepackage{mathrsfs}  
\usepackage{hyperref}

\DeclareMathOperator*{\argmax}{\arg\!\max}
\usepackage[ruled,vlined]{algorithm2e}
\usepackage[inline]{enumitem}

\begin{document}
\title{A Deep Reinforcement Learning Framework for Eco-driving in Connected and Automated Hybrid Electric Vehicles}

\author{Zhaoxuan~Zhu,~\IEEEmembership{Member,~IEEE,}
        Shobhit~Gupta,~\IEEEmembership{Member,~IEEE,}
        Abhishek~Gupta,~\IEEEmembership{Member,~IEEE}
        Marcello~Canova,~\IEEEmembership{Member,~IEEE}
\thanks{Copyright (c) 2015 IEEE. Personal use of this material is permitted. However, permission to use this material for any other purposes must be obtained from the IEEE by sending a request to pubs-permissions@ieee.org.}
\thanks{Z. Zhu is with Motional. The work is done when he is with Center for Automotive Research, The Ohio State University, Columbus, OH, 43212 USA. Email: zhu.1083@osu.edu}
\thanks{S. Gupta is with General Motors Research \& Development. The work is done when he is with Center for Automotive Research, The Ohio State University. Email: gupta.852@buckeyemail.osu.edu}
\thanks{M. Canova is with Center for Automotive Research, The Ohio State University. Email: canova.1@osu.edu}
\thanks{A. Gupta is with Department of
Electrical and Computer Engineering, The Ohio State University, Columbus, OH, 43210 USA. Email: gupta.706@osu.edu}
}

\markboth{IEEE TRANSACTIONS ON Vehicular Technology}%
{Zhu \MakeLowercase{\textit{et al.}}: A Deep Reinforcement Learning Framework for Eco-driving in Connected and Automated Hybrid Electric Vehicles}

\maketitle

\begin{abstract}
Connected and Automated Vehicles (CAVs), in particular those with multiple power sources, have the potential to significantly reduce fuel consumption and travel time in real-world driving conditions.
In particular, the eco-driving problem seeks to design optimal speed and power usage profiles based upon look-ahead information from connectivity and advanced mapping features, to minimize the fuel consumption over a given itinerary. 

In this work, the eco-driving problem is formulated as a Partially Observable Markov Decision Process (POMDP), which is then solved with a state-of-art Deep Reinforcement Learning (DRL) Actor Critic algorithm, Proximal Policy Optimization.
An eco-driving simulation environment is developed for training and evaluation purposes.
To benchmark the performance of the DRL controller, a baseline controller representing the human driver, a trajectory optimization algorithm and the wait-and-see deterministic optimal solution are presented.
With a minimal onboard computational requirement and a comparable travel time, the DRL controller reduces the fuel consumption by more than 17\% compared against the baseline controller by modulating the vehicle velocity over the route and performing energy-efficient approach and departure at signalized intersections, over-performing the more computationally demanding trajectory optimization method.
\end{abstract}

\begin{IEEEkeywords}
Connected and automated vehicle, eco-driving, deep reinforcement learning, dynamic programming, long short-term memory.
\end{IEEEkeywords}

\IEEEpeerreviewmaketitle

\section{Introduction}
\IEEEPARstart{W}ith the advancement in the vehicular connectivity and autonomy, Connected and Automated Vehicles (CAVs) have the potential to operate in a more time- and fuel-efficient manner \cite{vahidi2018energy}.
With Vehicle-to-Vehicle (V2V) and Vehicle-to-Infrastructure (V2I) communication, the controller has access to real-time look-ahead information including the terrain, infrastructure and surrounding vehicles.
Intuitively, with connectivity technologies, controllers can plan a speed profile that allows the ego vehicle to intelligently pass more signalized intersections in green phases with less change in speed.
This problem is formulated as the eco-driving problem (incorporating Eco-Approach and Departure at signalized intersections), which aims to minimize the fuel consumption and the travel time between two designated locations by co-optimizing the speed trajectory and the powertrain control strategy \cite{sciarretta2015optimal, jin2016power}.

The literature related to the eco-driving problem distinguishes among two aspects, namely, powertrain configurations and traffic scenarios.
Regarding powertrain configuration, the difference is in whether the powertrain is equipped with a single power source \cite{ozatay2014cloud, jin2016power, han2019fundamentals, sun2020optimal} or a hybrid electric architecture \cite{mensing2012vehicle, guo2016optimal, olin2019reducing, bae2019real}.
The latter involves modeling multiple power sources and devising optimal control algorithms that can synergistically split the power demand to efficiently utilize the electric energy stored in the battery.
Maamria et al. \cite{maamria2018computation} systematically compare the computational requirement and the optimality of different eco-driving formulations solved offline via Deterministic Dynamic Programming (DDP).

Related to the traffic scenarios, Ozatay et al. \cite{ozatay2014cloud} proposed a framework providing advisory speed profile using online optimization conducted on a cloud-based server without considering the real-time traffic light variability.
Olin et al. \cite{olin2019reducing} implemented the eco-driving framework to evaluate real-world fuel economy benefits obtained from a control logic computed in a Rapid Prototyping System on-board a test vehicle.
As traffic lights are not explicitly considered in these studies, the eco-driving control module is required to be coupled with other decision-making agents, such as human drivers or Adaptive Cruise Control (ACC) systems.
Other studies have explicitly modeled and considered Signal Phase and Timings (SPaTs).
Jin et al. \cite{jin2016power} formulated the problem as a Mixed Integer Linear Programming (MILP) for conventional vehicles with Internal Combustion Engine (ICE).
Asadi et al. \cite{asadi2011predictive} used traffic simulation models and proposed to solve the problem considering probabilistic SPaT with DDP.
Sun et al. \cite{sun2020optimal} formulated the eco-driving problem as a distributionally robust stochastic optimization problem with collected real-world data.
Guo et al. \cite{guo2016optimal} proposed a bi-level control framework with a hybrid vehicle.
Bae \cite{bae2019real} extended the work in \cite{sun2020optimal} to a heuristic HEV supervisory controller.
Deshpande et al. \cite{deshpande2021real,deshpande2021vehicle} designed a hierarchical Model Predictive Control (MPC) strategy to combine a heuristic logic with intersection passing capability with the strategy in \cite{olin2019reducing}.
Guo \cite{guo2018fuel,guo2018pmp} developed hierarchical planning and control modules for the fuel efficiency in vehicle platooning, and later on extended the work to vehicle fleets \cite{guo2022distributed1,guo2022distributed2}.

The dimensionality of the problem, therefore the computational requirements, can become quickly intractable as the number of states increases.
This is the case, for instance, when the energy management of a hybrid powertrain system is combined with velocity optimization in presence of other vehicles or approaching signalized intersections.
The aforementioned methods either consider a simplified powertrain model\cite{mensing2013trajectory, bae2019real} or treat the speed planning and the powertrain control hierarchically \cite{guo2016optimal}.
Although such efforts made the real-time implementation feasible, the optimality can be sacrificed \cite{maamria2018computation}.
The use of Deep Reinforcement Learning (DRL) in the context of eco-driving has caught considerable attention in recent years.
DRL provides a train-offline, execute-online methodology with which the policy is learned from historical data or by interacting with a simulated environment.
The offline training can either result in an explicit policy or be implemented as part of MPC as the terminal cost function.
Shi et al. \cite{shi2018application} modeled the conventional vehicles with ICE as a simplified model and implemented Q-learning to minimize the $CO_2$ emission at signalized intersections.
Li et al. \cite{li2019ecological} apply an actor-critic algorithm on the ecological ACC problem in car-following mode.
Pozzi et al. \cite{pozzi2020ecological} designed a velocity planner with Deep Deterministic Policy Gradient (DDPG) that operates on top of an ACC module for safety concerns and considers the signalized intersection and hybrid powertrain configuration.

This work focuses on the development of the eco-driving controller for HEVs with the capability to pass signalized intersections autonomously under urban and highway conditions.
The contribution of this work is threefold. 
\begin{enumerate*} [label=\itshape\alph*\upshape)]
\item Compared to the previous applications of DRL on the eco-driving problem \cite{li2019ecological, shi2018application}, the eco-driving problem is formulated as a centralized problem where a physics-based quasi-static nonlinear hybrid electric powertrain model is considered.
\item To overcome the intensive onboard computation, a novel Partially Observable Markov Decision Process (POMDP) eco-driving formulation is proposed and subsequently solved with an actor-critic DRL algorithm, Proximal Policy Optimization (PPO), along with Long Short-Term Memory (LSTM) as the function approximators.  
In addition, the design of the reward mechanism, particularly regarding the behaviors at signalized intersections, is discussed in detail. 
\item A co-simulation framework that integrated powertrain model and traffic simulation is proposed such that routes directly sampled from city map on large scale can be used for training and evaluation.
\end{enumerate*}
Compared to the aforementioned studies, this paper demonstrates that, by using modern DRL algorithm along with an efficient and high-fidelity environment model, a statistical superior policy with reasonable onboard computation can be learned for the task of eco-driving for HEVs under real-life driving routes, which is very challenging to tackle onboard by classical approaches.

To benchmark the performance of the resultant explicit policy, we present a baseline strategy representing human driving behaviors with a rule-based energy management module, a trajectory optimization strategy in \cite{deshpande2021vehicle,deshpande2021real} and a wait-and-see deterministic optimal solution.
The comparison was conducted over 100 randomly generated trips with Urban and Mixed Urban driving scenarios.

The remainder of the paper is organized as follows.
Section \ref{sec: env_mdl} presents the simulation environment.
Section \ref{sec: drl_preliminaries} introduces the preliminaries of the DRL algorithm employed in this work.
Section \ref{sec: formulation} mathematically formulates the eco-driving problem, and Section \ref{sec: drl_adoption} presents the proposed DRL controller.
Section \ref{sec: methods for benchmarking} presents the strategies used for benchmarking.
Section \ref{sec: results} shows the training details and benchmarks the performance.

\section{Environment Model}\label{sec: env_mdl}
\begin{figure}[]
    \centering
    \includegraphics[width=\columnwidth]{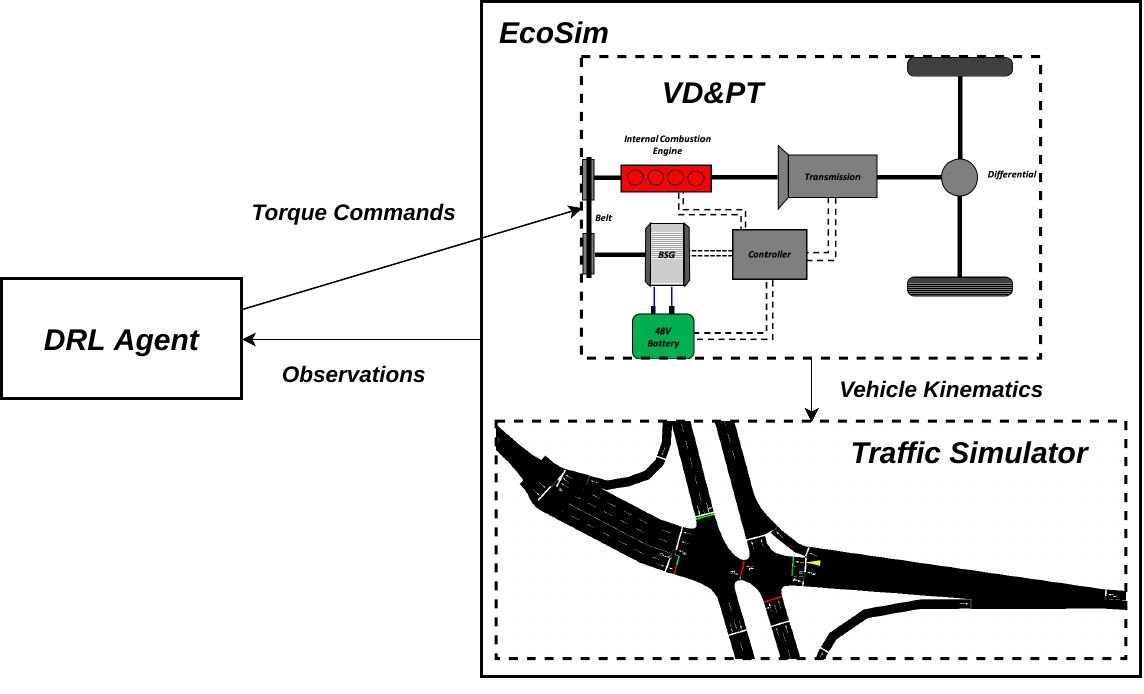}
    \caption{The Structure of The Environment Model}
    \label{fig: environment}
\end{figure}

The successful training of the reinforcement learning agent relies on an environment to provide the data.
In particular, model-free reinforcement learning methods typically require a large amount of data before agents learn the policy via the interaction with the environment.
In the context of eco-driving, collecting such an amount of real-world driving data is expensive.
Furthermore, the need for policy exploration during training poses safety concerns for human operators and hardware.
For these reasons, a model of the environment is developed for training and validation purposes.
The environment model, demonstrated in Fig.\ref{fig: environment}, consists of a Vehicle Dynamics and Powertrain (VD\&PT) model and a microscopic traffic simulator.
The environment model is discretized with time difference of 1 $s$.
The controller commands three control inputs, namely, the ICE torque, the electric motor torque and the mechanical brake torque.
The component-level torques collectively determine the HEV powertrain dynamics, the longitudinal dynamics of the ego vehicle and its location along the trip.
While the states of vehicle and powertrain such as battery State-of-Charge ($SoC$), velocity and gear are readily available to the powertrain controller, the availability of the connectivity information depends on the infrastructure and the types of sensors equipped onboard.
In this study, it is assumed that Dedicated Short Range Communication (DSRC) \cite{asadi2010predictive} sensors are available onboard, and SPaT becomes available and remains accurate once it enters the 200 $m$ range.
The uncertainties caused by sensor unavailability and inaccuracy in SPaT, as studied in \cite{sun2020optimal,mahler2014optimal} in SPaT, is not considered in the simulation model or in the study.
While adding uncertainties in the traffic model is left as future work, such uncertainties can be ingested by the Markov Decision Process (MDP) formulation. 
Thus, the model-free DRL problem formulation is expected to remain the same.
The DRL agent utilizes the SPaT from the upcoming traffic light while ignoring the SPaT from any other traffic light regardless of the availability.
Specifically, the distance to the upcoming traffic light, its status and SPaT program are fed into the controller as observations.
Finally, a navigation application with Global Positioning System (GPS) is assumed to be on the vehicle such that the locations of the origin and the destination, the remaining distance, the speed limits of the entire trip are available at every point during the trip.

\subsection{Vehicle and Powertrain Model} \label{sec: veh_dynamisc}
\begin{figure}[] 
	\centering
	\includegraphics[width=\columnwidth]{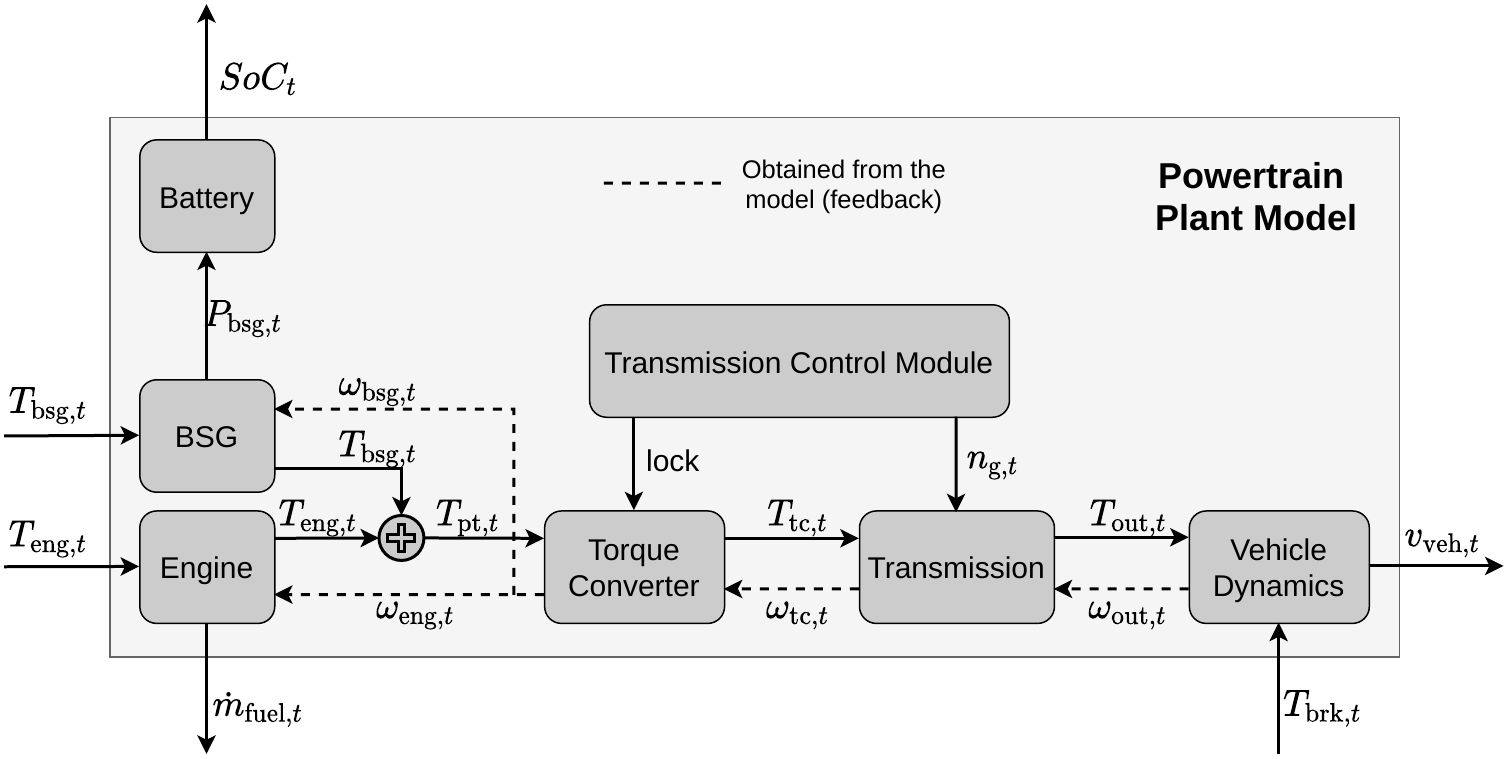}
	\caption{Block Diagram of 48V P0 Mild-Hybrid Drivetrain.}
	\label{fig_plant_model}
\end{figure}

A forward-looking dynamic powertrain model is developed for fuel economy evaluation and control strategy verification over real-world routes. In this work, a P0 mild-hybrid electric vehicle (mHEV) is considered, equipped with a 48V Belted Starter Generator (BSG) performing torque assist, regenerative braking and start-stop functions. The diagram of the powertrain is illustrated in Fig.\ref{fig_plant_model}. The key components of the low-frequency quasi-static model are described below.

\subsubsection{Engine Model} \label{Engine Model Description}
The engine is modeled as low-frequency quasi-static nonlinear maps. The fuel consumption and the torque limit maps are based on steady-state engine test bench data provided by a supplier:
\begin{equation}
\dot{m}_{\text{fuel},t}=\psi\left(\omega_{\text{eng},t},T_{\text{eng},t}\right),
\label{eq_T_eng}
\end{equation}
where the second subscript $t$ represents the discrete time index, and $\omega_{\text{eng}}$ and $T_{\text{eng}}$ are the engine angular velocity and torque, respectively.

\subsubsection{BSG Model} \label{BSG Model Description}
In a P0 configuration, the BSG is conneted to the engine via a belt, as shown in Eqn. \ref{eq_bsg_belt_ratio}. A simplified, quasi-static efficiency map ($\eta(\omega_{\text{bsg},t},T_{\text{bsg},t})$) is used to compute the electrical power output $P_{\text{bsg},t}$ in both regenerative braking and traction operating modes:
\begin{gather}
	\omega_{\text{bsg},t} = \tau_{\text{belt}}   \omega_{\text{eng},t},  \label{eq_bsg_belt_ratio}\\
	P_{\text{bsg},t} = T_{\text{bsg},t}   \omega_{\text{bsg},t}    \begin{cases}
	\eta(\omega_{\text{bsg},t},T_{\text{bsg},t}), & T_{\text{bsg},t}<0  \\
	\frac{1}{\eta(\omega_{\text{bsg},t},T_{\text{bsg},t})}, & T_{\text{bsg},t}>0  \\	    
	\end{cases}		\label{eq_P_bsg}
\end{gather}
where $\tau_{\text{belt}}$, $\omega_{\text{bsg},t}$ and $T_{\text{bsg},t}$ refer to the belt ratio, the BSG angular velocity and the BSG torque, respectively.

\subsubsection{Battery Model} \label{Battery Model Description}
A zero-th order equivalent circuit model is used to model the current ($I_t$) dynamics. Coulomb counting \cite{rong2006analytical} is used to compute the battery $SoC$:
\begin{gather}
	I_t = \frac{V_{\text{OC}}(SoC_t) - \sqrt{V_{\text{OC}}^2(SoC_t) -4 R_0(SoC_t) P_{\text{bsg},t}}}{2R_0(SoC_t)}, \label{eq_I_batt}\\
	SoC_{t+1} = SoC_t - \frac{\Delta t}{C_{\text{nom}}} (I_t + I_{\text{a}}),
\end{gather}
where $\Delta_t$ is the time discretization, which is set to 1s in this study. The power consumed by the auxiliaries is modeled by a calibrated constant current bias $I_{\text{a}}$. The cell open circuit voltage $V_{\text{OC}} (SoC_t)$ and internal resistance $R_0 (SoC_t)$ data are obtained from the pack supplier. 

\subsubsection{Torque Converter Model} \label{T/C Model Description}
A simplified torque converter model is developed to compute the losses during traction and regeneration modes. Here, the lock-up clutch is assumed to be always actuated, applying a controlled slip $\omega_{\text{slip}}$ between the turbine and the pump. The assumption might be inaccurate during launches, and this can be compensated by including a fuel consumption penalty in the optimization problem, associated to each vehicle launch event. This model is described as follows \cite{livshiz2004validation}:
\begin{gather}
	T_{\text{tc},t} = T_{\text{pt},t},  \label{eq_T_t}\\
	\omega_{\text{p},t} = \omega_{\text{tc},t} + \omega_{\text{slip}}\left(n_{\text{g},t},\omega_{\text{eng},t},T_{\text{eng},t}\right) \label{eq_w_p},\\
	\omega_{\text{eng},t} = \begin{cases}
		\omega_{\text{p},t}, & \omega_{\text{p},t} \geq \omega_{\text{stall}} \\
		\omega_{\text{idle},t}, & 0\leq \omega_{\text{p},t} < \omega_{\text{stall}} \\
		0, & 0\leq \omega_{\text{p},t} < \omega_{\text{stall}} \text{ and $\text{Stop} = 1$}
	\end{cases} \label{eq_w_eng} 
\end{gather}
where $n_{\text{g}}$ is the gear number, $\omega_{\text{p},t}$ is the speed of the torque converter pump, $\omega_{\text{tc},t}$ is the speed of the turbine, $\omega_{\text{stall}}$ is the speed at which the engine stalls, $\omega_{\text{idle},t}$ is the idle speed of the engine, stop is a flag from the ECU indicating engine shut-off when the vehicle is stationary, $T_{\text{tc},t}$ is the turbine torque, and $T_{\text{pt},t}$ is the combined powertrain torque. The desired slip $\omega_{\text{slip}}$ is determined based on the powertrain conditions and desired operating mode of the engine (traction or deceleration fuel cut-off).

\subsubsection{Transmission Model} \label{Transmission Model Description}
The transmission model is based on a static gearbox, whose equations are as follows:
\begin{gather}
    \begin{split}
    \omega_{\text{tc},t} &= \tau_{\text{g}}(n_{\text{g},t})   \omega_{\text{trans},t} \\ 
    &= \tau_{\text{g}}(n_{\text{g},t})   \tau_{\text{fdr}}  \omega_{\text{out},t}\\
    &= \tau_{\text{g}}(n_{\text{g},t})   \tau_{\text{fdr}}   \frac{v_{\text{veh},t}}{R_w}, \label{eq_w_t}
    \end{split}\\
    T_{\text{trans},t} = \tau_{\text{g}}(n_{\text{g},t})   T_{\text{tc},t}, \\
	T_{\text{out},t} = \begin{cases}
	 \tau_{\text{fdr}} \eta_{\text{trans}}(n_{\text{g},t}, T_{\text{trans},t}, \omega_{\text{trans},t})  T_{\text{trans},t}, & T_{\text{trans},t} \geq 0 \\
		\dfrac{ \tau_{\text{fdr}} }{\eta_{\text{trans}}(n_{\text{g},t}, T_{\text{trans},t}, \omega_{\text{trans},t})}  T_{\text{trans},t}, & T_{\text{trans},t} < 0
	\end{cases} \label{eq_T_out}
\end{gather}
where $\tau_{\text{g}}$ and $\tau_{\text{fdr}}$ are the gear ratio and the final drive ratio, respectively. The transmission efficiency $\eta_{\text{trans}}(n_\text{g}, T_{\text{trans}}, \omega_{\text{trans}})$ is scheduled as a nonlinear map expressed as a function of gear number $n_\text{g}$, transmission input shaft torque $T_{\text{trans},t}$ and transmission input speed $\omega_{\text{trans}}$. $\omega_{\text{out},t}$ refers to the angular velocity of the wheels. $R_w$ and $v_{\text{veh},t}$ are the radius of the vehicle wheel and the longitudinal velocity of the vehicle, respectively.

\subsubsection{Vehicle Longitudinal Dynamics Model} \label{Road Load Model Description}
The vehicle dynamics model is based on the road-load equation, which indicates the tire rolling resistance, road grade, and aerodynamic drag:
\begin{gather}
\begin{split}
	a_{\text{veh},t} = \frac{T_{\text{out},t}-T_{\text{brk},t}}{M R_\text{w}} &-\dfrac{1}{2}  \dfrac{C_\text{d} \rho_\text{a}   A_\text{f}}{M}   v_{\text{veh},t}^2 \\
	&-  g   C_\text{r}   \cos{\alpha_t}   v_{\text{veh},t} -  g  \sin{\alpha_t}. \label{eq_V_veh_state}
\end{split}
\end{gather}
Here, $a_{\text{veh},t}$ is the longitudinal acceleration of the vehicle, $T_{\text{brk}}$ is the brake torque applied on wheel, $M$ is the mass of the vehicle, $C_{\text{d}}$ is the aerodynamic drag coefficient, $\rho_\text{a}$ is the air density, $A_{\text{f}}$ is the effective aerodynamic frontal area, $C_\text{r}$ is rolling resistance coefficient, and $\alpha_t$ is the road grade.

\subsubsection{Vehicle Model Verification} \label{Model Verification}
\begin{figure}[t!]
	\centering
	\includegraphics[width=\columnwidth]{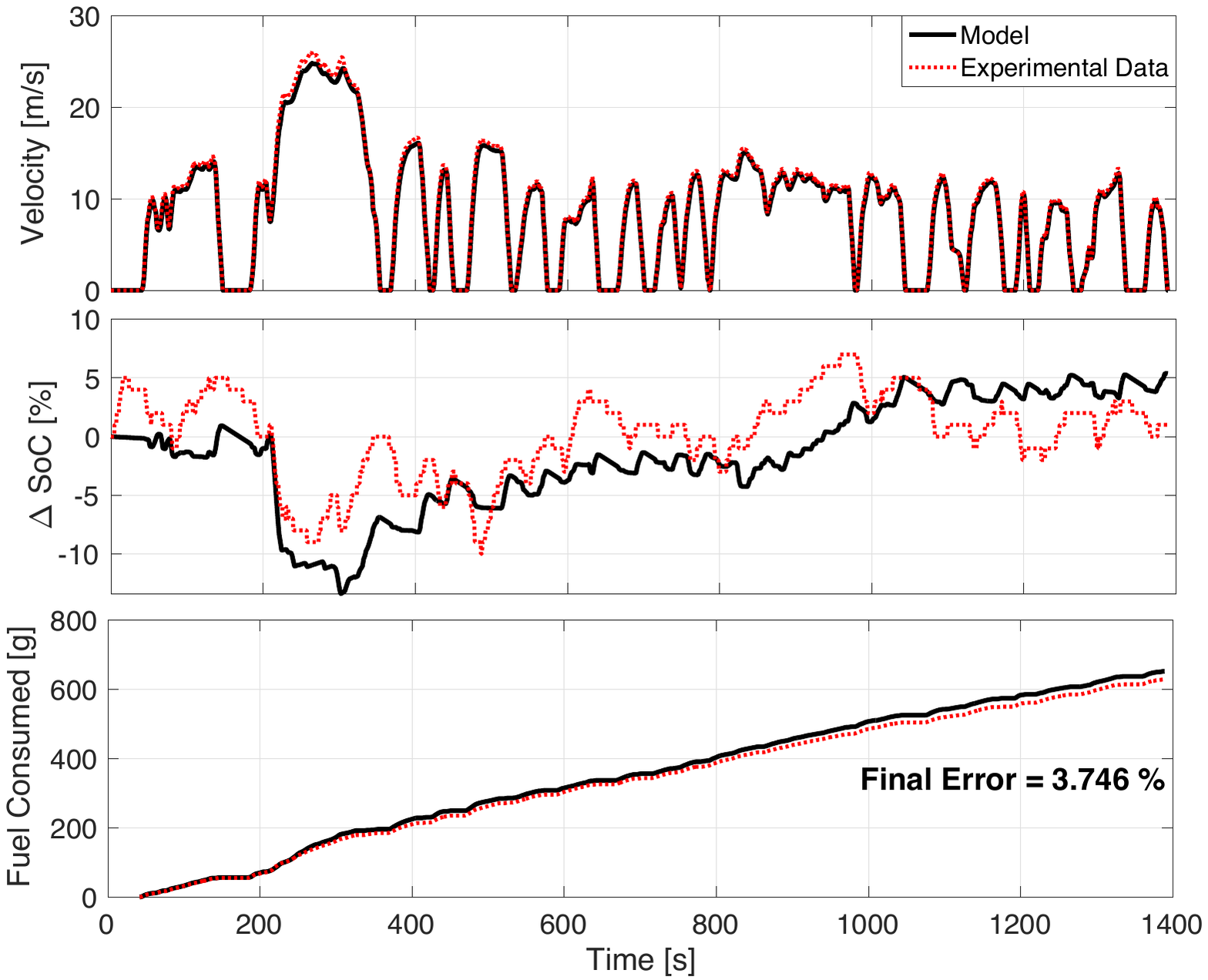}
	\caption{Validation of Vehicle Velocity, $SoC$ and Fuel Consumed over FTP Cycle.}
	\label{fig_veh_vel_soc_fuel_validation_FTP}
\end{figure}

The forward model is then calibrated and verified using experimental data from chassis dynamometer testing. The key variables used for evaluating the model are vehicle velocity, battery $SoC$, gear number, engine speed, desired engine and BSG torque profiles, and fuel consumption. Fig. \ref{fig_veh_vel_soc_fuel_validation_FTP} show sample results from model verification over the Federal Test Procedure (FTP) regulatory drive cycle, where the battery $SoC$ and fuel consumption are compared against experimental data. 

The mismatches in the battery $SoC$ profiles can be attributed to the simplicity of the battery model, in which the electrical accessory loads are modeled using a constant current bias.
The fuel consumption over the FTP cycle is well estimated by the model, with an error on the final value less than 4\% relative to the experimental data.

\subsection{Traffic Model}
A large-scale microscopic traffic simulator is developed in the open source software Simulation of Urban Mobility (SUMO) \cite{SUMO2018} as part of the environment.
To recreate realistic mixed urban and highway trips for training, the map of the city of Columbus, OH, USA is downloaded from the online database OpenStreetMap \cite{OpenStreetMap}.
The map contains the length, shape, type and speed limit of the road segments and the detailed program of each traffic light in signalized intersections.

\begin{figure}[!t]
    \centering
    \includegraphics[width=\columnwidth]{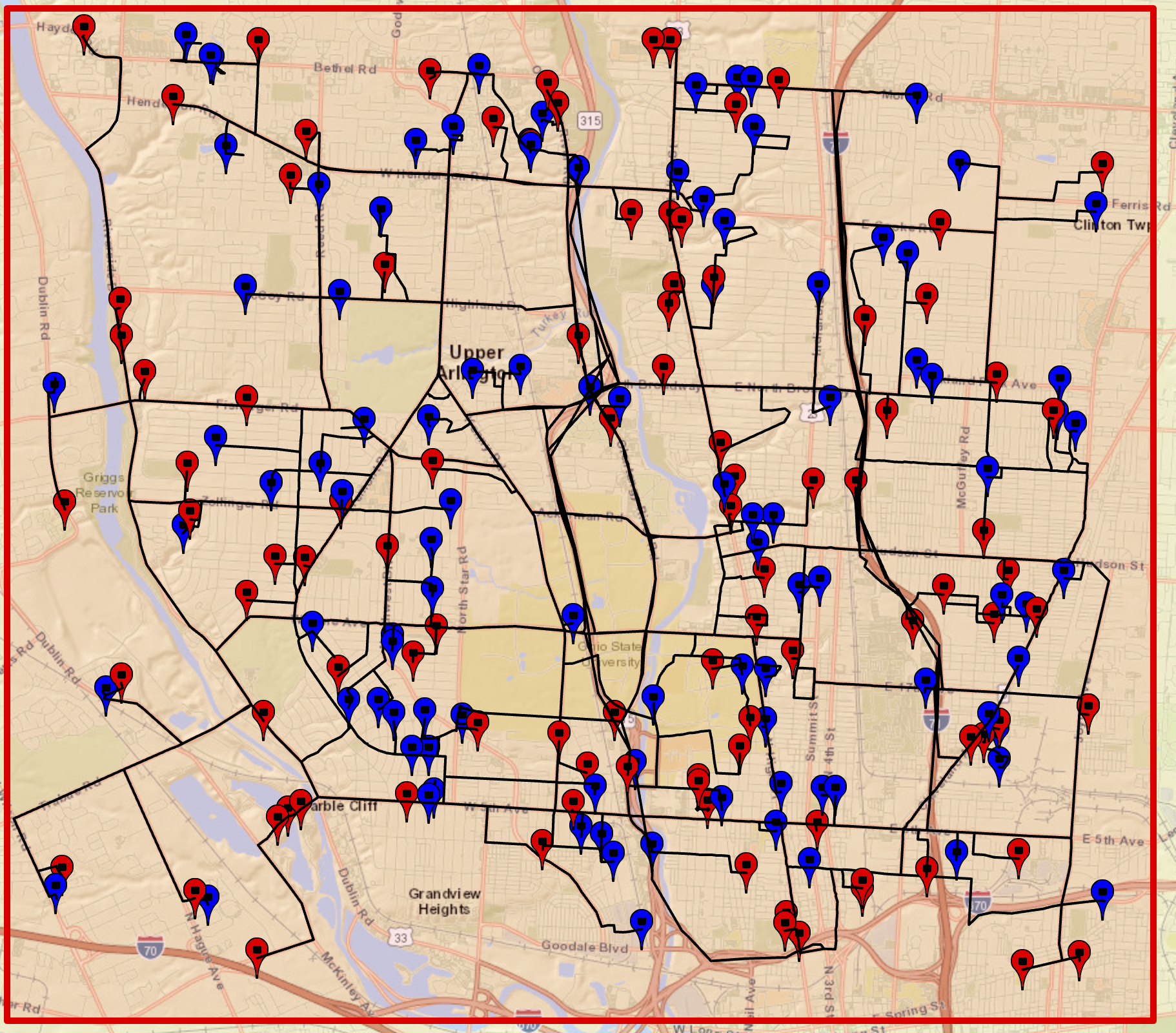}
    \caption{Map of Columbus, OH for DRL Training [Each red and blue marker denotes the start and end point of an individual trip and the colored line denote the route between these points.]}
    \label{fig:map_of_columbus}
\end{figure}

Fig. \ref{fig:map_of_columbus} highlights the area ($\sim 11 km$ by $11 km$) covered in the study.
In the area, 10,000 random passenger car trips are generated as the training set, and the total distance of each trip is randomly distributed from 5 $km$ to 10 $km$.
Another 100 trips, with which the origins and the destinations are marked in red and blue in Fig. \ref{fig:map_of_columbus}, respectively, are generated following the same distribution as the testing set.
In addition, the inter-departure time of each trip follows a geometric distribution with the success rate $p=0.01$.
The variation and the randomness of the trips used for training enhance the richness of the environment, which subsequently leads to a learned policy that is less subject to local minima and agnostic to specific driving conditions (better generalizability) \cite{heess2017emergence}.

The interface between the traffic simulator and the VD\&PT model is established via Traffic Control Interface (TraCI) as part of the SUMO package. At any given time step, the kinetics of the vehicle calculated from VD\&PT is fed to the traffic simulator as input. Subsequently, SUMO determines the location of the ego vehicle, updates the connectivity information such as the SPaT of the upcoming traffic light and the GPS signal and returns them to the agent as part of the observations. 

\section{Deep Reinforcement Learning Preliminaries}\label{sec: drl_preliminaries}
\subsection{Markov Decision Process}
In Markov Decision Process (MDP), sequential decisions are made in order to maximize the discounted sum of the rewards. An MDP can be defined by a tuple $\langle\mathcal{S},\mathcal{A},P,\rho_0,r,\gamma\rangle$, where $\mathcal{S}$ and $\mathcal{A}$ are the state space and the action space, respectively; $P:\mathcal{S}\times \mathcal{A} \times \mathcal{S} \rightarrow [0, 1]$ is the transition dynamics distribution; $\rho_0$ is the initial distribution of the state space. The reward function $r:\mathcal{S}\times\mathcal{A}\times\mathcal{S}\rightarrow \mathbb{R}$ is a function that maps the tuple $\left(s_t,a_t,s_{t+1} \right)$ to instantaneous reward. Finally, $\gamma$ is the discount factor that prioritizes the immediate reward and ensures that the summation over infinite horizon is finite.

Let $\pi:\mathcal{S}\times\mathcal{A} \rightarrow [0, 1]$ be a randomized policy and $\Pi$ be the set of all randomized policies. The objective of MDP is to find the optimal policy $\pi^*$ that minimizes the expectation of the discounted sum of the rewards defined as follows:
\begin{equation}
\begin{gathered}
    \pi^* = \argmax_{\pi \in \Pi}\eta(\pi), \; \text{where} \\
    \eta(\pi) = \mathbb{E}_{s_{t+1}\sim \mathcal P(\cdot|s_t,a_t)}\left[ \sum_{t=0}^\infty  \gamma^{t}r\left(s_t,a_t\right)\right],  \\
    \text{where } s_0 \sim \rho_0(\cdot), \quad a_t\sim\pi(\cdot|s_t).
\end{gathered}
\end{equation}
In the remaining work, the expectation under the state trajectory $E_{s_{t+1}\sim P(\cdot|s_t,a_t)}[ \cdot ]$ will be written compactly as $E_{\pi}[\cdot ]$. For any policy $\pi$, the value function $V^\pi:\mathcal{S}\rightarrow \mathbb{R}$, the Q function $Q^\pi:\mathcal{S}\times \mathcal{A} \rightarrow \mathbb{R}$ and the advantage function $A^\pi:\mathcal{S}\times \mathcal{A} \rightarrow \mathbb{R}$ are defined as follows:
\begin{gather}
    V^{\pi}(s_t) = \mathbb{E}_{\pi}\left[\sum_{i=t}^\infty\gamma^{i-t}r\left(s_i,a_i\right)|s_t\right], \\
    Q^{\pi}(s_t,a_t) =\mathbb{E}_{\pi}\left[\sum_{i=t}^\infty\gamma^{i-t}r\left(s_i,a_i\right)|s_t,a_t\right], \\
    A^{\pi}(s_t,a_t) =Q^{\pi}(s_t,a_t) - V^{\pi}(s_t).
\end{gather}

\subsection{Actor-critic Algorithm}
The actor-critic algorithm is one of the earliest concepts in the field of reinforcement learning \cite{barto1983neuronlike, sutton2018reinforcement}. The actor is typically a stochastic control policy, whereas the critic is a value function assisting the actor to improve the policy. For DRL, both the actor and critic are typically in the form of deep neural networks.     

In this study, policy gradient method is used to iteratively improve the policy. According to Policy Gradient Theorem in \cite{williams1992simple ,marbach2001simulation}, the gradient of the policy, parameterized by $\theta$, can be determined by the following equation:
\begin{align}
    \nabla_\theta \eta(\pi_{\theta_k}) \propto \sum_s \rho(s) \sum_a Q^{\pi_{\theta_k}}(s,a)\nabla_\theta \pi_{\theta_k}(a|s),
\end{align}
where $\rho(s)$ is the discounted on-policy state distribution defined as follows:
\begin{align}
    \rho(s) = \sum_{t=0}^\infty\gamma^tP(s_t=s).
\end{align}
As in \cite{sutton2000policy}, the gradient can be estimated as follows:
\begin{align}
     \nabla_\theta \eta(\pi_{\theta_k}) = \mathbb{E}_{\pi}\left[ A^{\pi_{\theta_k}}(s_t,a_t) \dfrac{\nabla_\theta \pi_{\theta_k}(a_t|s_t)}{\pi_{\theta_k}(a_t|s_t)} \right]. \label{eq:grad estimation}
\end{align}
Accordingly, to incrementally increase $\eta(\pi_\theta)$, the gradient ascent rule follows
\begin{align}
    \theta_{k+1} = \theta_k+\alpha_k\nabla_{\theta}\eta(\pi_{\theta_k}),
\end{align}
where $\alpha_k$ is the learning rate.
As Eqn. \eqref{eq:grad estimation} is a local estimation of the gradient at the neighborhood of the current policy, updating the policy in such a direction with a large step size could potentially lead to a large performance drop.
Schulman et al. \cite{schulman2015trust} proposed to constrain the difference between the probability distributions of the old and the new policy with the trust region method.
Although being less brittle, the algorithm requires the analytical Hessian matrix, resulting in a high computational load and a nontrivial implementation.
In this paper, a first-order method proposed by Schulman et al. \cite{schulman2017proximal} is used.
Instead of Eqn. \eqref{eq:grad estimation}, a clipped surrogate objective function is defined as follows:
\begin{equation}
\begin{gathered}
\begin{split}
L_t(\theta) = \mathbb{E}_{\pi} &\bigl[ \min \left( r_t(\theta), \right. \bigr.\\
 &\quad \quad \: \bigl. \left. \text{clip}\left( r_t(\theta),1-\epsilon, 1+\epsilon \right)\right)A^{\pi_{\theta_k}}(s_t,a_t)\bigr],\end{split} \\
\text{where} \; r_t(\theta) = \dfrac{\pi_\theta(a_t|s_t)}{\pi_{\theta_k}(a_t|s_t)},\\
\text{clip}(x, a_\text{min}, a_\text{max})=\min\left(\max\left(x, a_\text{min}\right), a_\text{max}\right).\label{eq: ppo_update_rule}
\end{gathered}
\end{equation}
Here, the hyperparameter $\epsilon$ is the clipping ratio.
Note the first-order derivative of the loss function around $\theta_k$, $\nabla_\theta L_t(\theta)|_{\theta_k}$, is equal to $\nabla_\theta \eta(\pi_{\theta})|_{\theta_k}$ which is consistent with Policy Gradient Theorem.


\subsection{Partially Observable Markov Decision Process}
In many practical applications, states are not fully observable. The partial observability can arise from sources such as the need to remember the states in history, the sensor availability or noise and unobserved variations of the plant under control \cite{heess2015memory}. Such a problem can be modeled as a POMDP where observations $o_t \in \mathcal{O}$, instead of states, are available to the agent. The observation at certain time follows the conditional distribution given the current state $o_t \sim P(\cdot|s_t)$.

For POMDP, the optimal policy, in general, depends on the entire history $h_t=(o_1, a_1, o_2, a_2, ... a_{t-1},o_t)$, i.e. $a_t \sim \pi(\cdot|h_t)$.
The optimal policy can be obtained by giving policies the internal memory to access the history $h_t$.
In \cite{wierstra2007solving}, it is shown that the policy gradient theorem can be used to solve the POMDP problem with Recurrent Neural Network (RNN) as the function approximator, i.e. Recurrent Policy Gradients.
Compared to other function approximators such as Multilayer Perceptron (MLP) or Convolutional Neural Network (CNN), RNN exploits the sequential property of inputs and uses internal states for memory.
Specifically, Long Short-Term Memory (LSTM) \cite{hochreiter1997long}, as a special architecture of RNN, is typically used in DRL to avoid gradient explosion and gradient vanishing, which are the well-known issues to RNN \cite{bengio1994learning}.
In LSTM, three types of gates are used to keep the memory cells activated for arbitrarily long.
The combination of Policy Gradient and LSTM has shown excellent results in many modern DRL applications \cite{vinyals2019grandmaster, berner2019dota}.

In the eco-driving problem, there are trajectory constraints while approaching traffic lights and stop signs.
Thus, in these situations, the ego vehicle needs to remember the states visited in the recent past.
LSTM based function approximators are therefore chosen to approximate the value function and the advantage function so that the ego vehicle can use information about the past states visited to decide on its torques.

\section{Problem Formulation} \label{sec: formulation}
In the eco-driving problem, the objective is to minimize the weighted sum of fuel consumption and travel time between two designated locations. The optimal control problem (OCP) is formulated as follows:
\begin{subequations}
\begin{gather}
     \min_{T_{\text{eng}}, T_{\text{bsg}}, T_{\text{brk}}} \: \mathbb{E} \left[ \sum_{t=0}^\infty \left( \dot{m}_{\text{fuel},t} + c_{\text{time}} \right)   \Delta t \cdot \mathbb{I}\left[d_t < d_{\text{total}}\right] \right] \\
    \text{s.t.} \:  SoC_{t+1} = f_{\text{batt}}(v_{\text{veh},t}, SoC_t, T_{\text{eng},t}, T_{\text{bsg},t}, T_{\text{brk},t}) \\
     v_{\text{veh},t+1} = f_{\text{veh}}(v_{\text{veh}, t}, SoC_t, T_{\text{eng},t}, T_{\text{bsg},t}, T_{\text{brk},t}) \\
     T_{\text{eng}}^{\min}(\omega_{\text{eng},t}) \leq T_{\text{eng},t} \leq T_{\text{eng}}^{\max}(\omega_{\text{eng},t}) \label{eq:S2_con_Teng} \\
     T_{\text{bsg}}^{\min}(\omega_{\text{bsg},t}) \leq T_{\text{bsg},t} \leq T_{\text{bsg}}^{\max}(\omega_{\text{bsg},t}) \label{eq:S2_con_Tbsg} \\
     0 \leq T_{\text{brk},t} \leq T_{\text{brk}}^{\max} \label{eq:S2_con_Tbrk} \\
     I^{\min} \leq I_t \leq I^{\max} \label{eq:S2_con_I}\\
     \left\lbrace\left[\mathbb{I}(T_{\text{eng},t} > 0) \: \mathrm{OR} \: \mathbb{I}(T_{\mathrm{bsg},t} > 0)\right] \: \mathrm{AND} \: \mathbb{I}(T_{\mathrm{brk}, t} > 0)\right\rbrace \neq 0 \label{eq: torque_consistency_constraint}\\
     SoC^{\text{min}} \leq SoC_t \leq SoC^{\text{max}} \label{eq: soc_constraint}\\
    SoC_T \geq SoC_{\text{F}} \label{eq: terminal_soc_constraint}\\
    0 \leq v_{\text{veh},t} \leq v_{\text{lim},t} \label{eq: speed_limit}\\
    (t, d_t) \notin \mathcal{S}_{\text{red}} \label{eq: traffic_constraint}
\end{gather} \label{eq:OCP_formulation}
\end{subequations}

Here, $\dot{m}_{\text{fuel},t}$ is the instantaneous fuel consumption at time $t$, $c_\text{time}$ is a constant to penalize the travel time took at each step, $f_{\text{batt}}$ and $f_{\text{veh}}$ are the battery and vehicle dynamics introduced in Section \ref{sec: veh_dynamisc}.
The problem is formulated as an infinite horizon problem in which the stage cost becomes zero once the system reaches the goal, i.e. the traveled distance $d_t$ is greater than or equal to the total distance of the trip $d_{\text{total}}$.
Eqn. \eqref{eq:S2_con_Teng} to \eqref{eq: torque_consistency_constraint} are the constraints imposed by the powertrain components. 
Specifically, Eqn. \eqref{eq: torque_consistency_constraint} represents that the torques from the powertrain and the brake torque cannot be positive at the same time.
Eqn. \eqref{eq: soc_constraint} and Eqn. \eqref{eq: terminal_soc_constraint} are the constraints on the instantaneous battery $SoC$ and terminal $SoC$ for charge sustaining.
Here, the subscript $T$ represents the time at which the vehicle reaches the destination.
$SoC_\text{min}$, $SoC_\text{max}$ and $SoC_\text{F}$ are commonly set to 30\%, 80\% and 50\%.
Eqn. \eqref{eq: speed_limit} and \eqref{eq: traffic_constraint} are the constraints imposed by the traffic conditions.
The set $\mathcal{S}_{\text{red}}$ represents the set in which the traffic light at a certain location is in the red phase.

As the controller can only accurately predict the future driving condition in a relatively short-range due to the limited connectivity range and onboard processing power, the stochastic optimal control formulation is deployed to accommodate the future uncertainties. Specifically, since the surrounding vehicles are not considered in the study, the main source of uncertainty comes from the unknown SPaT and the road conditions such as the speed limits and the distance between signalized intersections beyond the connectivity range. 

\section{Deep Reinforcement Learning Controller} \label{sec: drl_adoption} 
\subsection{POMDP Adoption}
In this study, the eco-driving problem described by Eqn. \eqref{eq:OCP_formulation} is solved as a POMDP. The constraints on the action space, i.e. Eqn. \eqref{eq:S2_con_Teng}, \eqref{eq:S2_con_Tbsg} and \eqref{eq:S2_con_I}, are handled implicitly by a saturation function in the environment model, whereas the constraints on the state space are handled by imposing numerical penalties during the offline training.

\begin{table}[]
\centering
\caption{Observation and Action Space of the Eco-driving Problem}
\label{tab:state_action_space}
\begin{tabular}{c|c|p{5cm}}
& Variable & \multicolumn{1}{c}{Description}\\ [3pt]
\hline
\multirow{12}{*}{$\mathcal{O} \in \mathbb{R}^9 $} & $SoC$ & Battery $SoC$ \\[3pt]
& $v_{\text{veh}}$ & Vehicle velocity\\[3pt]
& $v_{\text{lim}}$ & Speed limit at the current road segment\\[3pt] 
& $v'_{\text{lim}}$ & Upcoming speed limit\\[3pt]
& $t_\text{s}$ & Time to the start of the next green light at the upcoming intersection\\[3pt] 
& $t_\text{e}$ & Time to the end of the next green light at the upcoming intersection\\[3pt] 
& $d_{\text{tfc}}$ & Distance to the upcoming traffic light\\[3pt]
& $d'_{\text{lim}}$ & Distance to the road segment of which the speed limit changes\\[3pt]
& $d_{\text{rem}}$ & Remaining distance of the trip\\[3pt]
\hline
\multirow{3}{*}{$\mathcal{A} \in \mathbb{R}^3 $} & $T_{\text{eng}}$ & Engine torque\\[3pt]
& $T_{\text{bsg}}$ & Motor torque \\[3pt]
& $T_{\text{brk}}$ & Equivalent brake torque
\end{tabular}
\end{table}

Tab. \ref{tab:state_action_space} lists the observation and action spaces used to approach the eco-driving POMDP.
Here, $SoC$ and $v_{\text{veh}}$ are the states measured by the onboard sensors, and $v_{\text{lim}}$, $v'_{\text{lim}}$, $d_{\text{tlc}}$, $d'_{\text{lim}}$ and $d_{\text{rem}}$ are assumed to be provided by the downloaded map and GPS.
$t_\text{s}$ and $t_\text{e}$ are the SPaT signals provided by V2I communication.
When the upcoming traffic light is in the green phase, $t_{\text{s}}$ remains $0$, and $t_{\text{e}}$ is the remaining time of the current green phase; when the upcoming traffic light is in red phase, $t_{\text{s}}$ is the remaining time of the current red phase, and $t_{\text{e}}$ is the sum of the remaining red phase and the duration of the upcoming green phase. 

Since the actor-critic algorithm is used in the paper, the action space is continuous.
The constraints in Eqn. \eqref{eq:S2_con_Teng} to \eqref{eq:S2_con_Tbrk} are incorporated by as satuation in the environment. 
Since the violation of the constraint in Eqn. \eqref{eq: torque_consistency_constraint} leads to the suboptimal strategy, i.e., the torques generated from the powertrain are wasted by the positive brake torque, no explicit constraint is applied w.r.t Eqn. \eqref{eq: torque_consistency_constraint}.

The reward function $r:\mathcal{S}\times\mathcal{A}\times\mathcal{S}\rightarrow\mathbb{R}$ consists of four terms:
\begin{align}
    r=\text{clip}\left[r_{\text{obj}}+r_{\text{vel}}+r_{\text{batt}}+r_{\text{tfc}}, -1, 1\right].
\end{align}
Here, $r_{\text{obj}}$ represents the rewards associated with the OCP objective; $r_{\text{vel}}$ is the penalty (negative reward) associated with the violation of the speed limit constraint Eqn. \eqref{eq: speed_limit}; $r_{\text{batt}}$ represents the penalties associated with the violation of the battery constraints imposed by Eqn. \eqref{eq: soc_constraint}, \eqref{eq: terminal_soc_constraint}; $r_{\text{tfc}}$ is the penalties regarding the violation of the traffic light constraint imposed by Eqn. \eqref{eq: traffic_constraint}. Specifically, the first three terms are designed as follows:
\begin{gather}
    r_{\text{obj},t} = c_{\text{obj}} \left( \dot{m}_{\text{fuel},t} [g/s] + c_{\text{time}} \right), \label{eq: objective_reward}\\
    r_{\text{vel},t} = c_{\text{vel,1}} \left[v_{\text{vel},t} - v_{\text{lim},t}\right]^+ + c_{\text{vel,2}} \dot{a}_{\text{veh},t}^2, \label{eq: speed_reward}\\
    r_{\text{batt},t} = \left( \begin{aligned}   &c_{\text{batt,1}} \left(\left[SoC_t - SoC^{\text{max}}\right]^+ \right.\\
    & \quad \quad \left. +\left[ SoC^{\text{min}} - SoC_t\right]^+ \right) , && d_t < d_{\text{total}} \\
    &c_{\text{batt},2} \left[ SoC_F - SoC_t \right]^+, && d_t \geq d_{\text{total}} \end{aligned} \right. \label{eq: battery_reward}
\end{gather}
where $[\cdot]^+$ is the positive part of the variable defined as
\begin{align}
    [x]^+ = \max(0, x).
\end{align}
In Eqn. \eqref{eq: speed_reward}, a penalty to the longitudinal jerk is assigned to the agent to improve the drive quality and to avoid unnecessary speed fluctuations. 

\begin{figure}[!t]
    \centering
    \includegraphics[width=1\columnwidth]{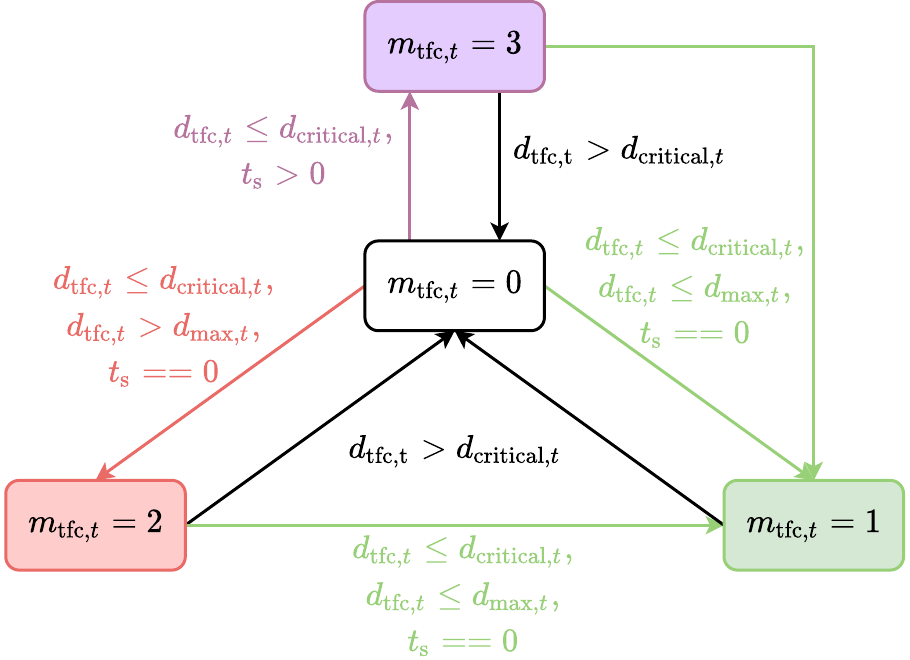}
    \caption{State Machine for the Indicator $m_{\text{tfc}}$}
    \label{fig:state_machine_driving_mode}
\end{figure}

While the design of the first three rewards is straightforward, the reward associated with the traffic light constraints is more convoluted to define. First, a discrete state variable $m_{\text{tfc}}$ is defined in Fig. \ref{fig:state_machine_driving_mode}. $m_{\text{tfc}}=0$ whenever the distance to the upcoming traffic light is greater than the critical braking distance $d_{\text{critical},t}$, which is defined as follows:
\begin{align}
    d_{\text{critical},t} = \dfrac{v_{\text{veh},t}^2}{2b_{\text{max}}},
\end{align}
where $b_\text{max}$ is the maximal deceleration of the vehicle. Intuitively, the agent does not need to make an immediate decision regarding whether to accelerate to pass or to decelerate to stop at the upcoming signalized intersection outside the critical braking distance range. Once the vehicle is within the critical braking distance range, $m_{\text{tfc}}$ is determined by the current status of the upcoming traffic light, the distance between the vehicle and the intersection and the maximal distance that the vehicle could drive given the remaining green phase $d_{\text{max},t}$ defined as follows:
\begin{gather}
    d_{\text{max},t} = \sum_{i=0}^{t_\text{e}}\left[\min\left( v_{\text{lim},t}, v_{\text{veh},t} + ia_{\text{max}} \right) \right],
\end{gather}
where $a_{\text{max}}$ is the maximal acceleration.

If the upcoming traffic light is in green, i.e. $t_s=0$, $m_{\text{tfc}}$ gets updated to $1$ if the distance between the vehicle and the upcoming intersection is less than the $d_{\text{max},t}$ and $2$ otherwise. Intuitively, $m_{\text{tfc}}=1$ means that the vehicle has enough time to cross the upcoming intersection within the remaining green phase in the current traffic light programming cycle. In case the vehicle following the actor policy was not able to catch the green light, a penalty proportional to $d_{\text{miss}}$, the distance between the vehicle at the last second of the green phase and the intersection, is assigned. On the other hand, $m_{\text{tfc}}=2$ means the vehicle would not reach the destination even with the highest acceleration in the current cycle. If the upcoming traffic light is in red phase, $m_{\text{tfc},t}$ gets updated to $3$. When the vehicle was not able to come to stop and violate the constraints, a penalty proportional to the speed at which the vehicle passes in red is assigned. As a summary, the reward associated with the traffic light constraints is designed as follows:
\begin{gather}
    r_{\text{tfc},t} = \begin{cases} c_{\text{tfc},1} + c_{\text{tfc},2} d_{\text{miss},t}, & m_{\text{tfc},t}=1\\c_{\text{tfc},1} + c_{\text{tfc},3} v_{\text{veh},t}, & \text{otherwise}
    \end{cases}
    \label{eq: traffic rewards}
\end{gather}

In Appendix \ref{sec: reward function design}, a guideline to the design and the tuning of the reward mechanism is provided, and the numerical values of all the constants in the reward function are listed in Table \ref{tab:values_for_constants}. In order to determine the reward at any given time, the environment model requires states that are not directly available as observations such as $\dot{a}_t$, $m_{\text{tfc}}$ and $d_{\text{miss}}$. Instead of making these states available to the control agent, the POMDP formulation is intentionally selected for two reasons. Firstly, $m_{\text{tfc}}$ is heuristically determined and ignores the acceleration limit imposed by the powertrain components. Since it poses a significant impact on the reward at the intersection, revealing $m_{\text{tfc}}$ to the controller results in a strong dependency of the strategy to it, and occasionally such a dependency misleads the agent to violate the constraints. Secondly, $d_{\text{miss}}$ is only relevant when the vehicle is catching a green light within the critical braking distance. Its numerical value in other situations could potentially mislead the policy, in the form of neural networks. 

Studies have suggested that clipping the rewards between $\left[-1, 1\right]$ results in a better overall training process \cite{mnih2015human, van2016learning}. With the coefficients listed in Table \ref{tab:values_for_constants}, the negative reward saturates at $-1$ when $d_{\text{miss},t}>75 m$ and $m_{\text{tfc},t}=1$ or $v_{\text{veh},t}>7.5 m/s$ and $m_{\text{tfc},1}=2$, which means that the rewarding mechanism would no longer differentiate the quality of the state-action pairs beyond these thresholds at the signalized intersection. Such a design, on one hand, reduces the strong impact of the heuristically designed $m_{\text{tfc}}$. On the other hand, it also significantly slows down, or in some cases, prevents any learning as the rewards carry little directional information. Heess et al. \cite{heess2017emergence} proposes to use a more diversified and rich environment to overcome the issue. In this study, the diversity of the environment is ensured by the size of the SUMO map and the 10,000 randomly generated trips. In addition, the vehicle speed $v_{\text{veh}}$ and battery $SoC$ are randomly assigned following $\mathcal{U}(0, v_{\text{lim}})$ and $\mathcal{U}(SoC^{\text{min}}, SoC^{\text{max}})$, respectively, at every $T=100$ time steps. This domain randomization mechanism, used in many other DRL applications \cite{zhu2020energy, berner2019dota}, forces the agent to explore the state space more efficiently and learn a more robust policy.

\subsection{Algorithm Details}
With the Monte Carlo method, the value function can be approximated as follows:
\begin{equation}
     \hat{V}^{\pi_{\theta}}_\xi(s_t) \leftarrow \hat{\mathbb{E}}_{\pi_{\theta}}\left[\sum_{i=t}^\infty\gamma^{i-t}r\left(s_i,a_i\right)|s_t\right],
\end{equation}
where the superscript $\pi_\theta$ indicates the value function is associated with the policy $\pi$ parameterized by $\theta$, and the subscript $\xi$ indicates the value function itself is parameterized by $\xi$. Although being unbiased, the Monte Carlo estimator is of high variance and requires the entire trajectory to be simulated. On the other hand, TD$(N)$ estimator is defined as follows:
\begin{equation}
    \hat{V}^{\pi_{\theta}}_\xi(s_t) \leftarrow \hat{\mathbb{E}}_{\pi_{\theta}}\left[\sum_{i=t}^{t+N}\gamma^{i-t}r\left(s_i,a_i\right) + \hat{V}^{\pi_{\theta}}_{\xi_{\text{old}}}(s_{t+N})|s_t\right].
\end{equation}
Compared to the Monte Carlo method, it reduces the required rollout length and the variance of the estimation by bootstrapping. However, TD(N) estimator is biased due to the approximation error in $V^{\pi}(s_{t+N})$. TD($\lambda$) included in \cite{sutton2018reinforcement} takes the geometric sum of the terms from TD(N), leading to an adjustable balance between bias and variance. In this study, LSTM is used as the function aproximators, instead of the data tuple $(s_t,a_t,r_t,s_{t+1})$, $(o_t, h_{o,t}, a_t, h_{a,t}, r_{t}, o_{t+1}, h_{o,t+1})$ are logged in simulation, where $h_{o}$ and $h_{a}$ are the hidden states of the policy and value function networks, respectively. Since the state space is randomized at every $N$ steps, truncated TD($\lambda$) is used for value function approximation. Specifically, after having collected the a sequence of tuples $\left( o_t, h_{o,t}, a_t, h_{a,t}, r_{t}, o_{t+1}, h_{o,t+1} \right)_{t_0:t_0+N}$, the following equations are used for updating the value function $\forall t \in [t_0, t_0+N-1]$:
\begin{gather}
\hat{V}^{\pi_{\theta}}_\xi(o_t,h_{o,t}) \leftarrow V^{\pi_\theta}_\xi(o_t,h_{o,t})+\sum_{i=t}^{t_0 + N - 1}(\gamma\lambda_V)^{i-t}\delta_i, \\ \label{eq value_function_update}
\text{where} \; \delta_i = r_i + \gamma V^{\pi_{\theta}}_{\xi}(o_{i+1}, h_{o,i+1})-V^{\pi_{\theta}}_{\xi}(o_i,h_{o,i}). 
\end{gather}

Similarly, to balance the variance and the bias, the advantage function estimation is estimated with truncated GAE$(\lambda)$ as proposed in \cite{schulman2015high}: 
\begin{gather}
    \hat{A}^{\pi_\theta}(o_t,h_{o,t},a_t) = \sum_{i=t}^{t_0+N-1} (\gamma\lambda_A)^{i-t}\delta_{i}. \label{eq: gae}
\end{gather}

\begin{figure}[]
	\centering
	\includegraphics[width=0.5\textwidth]{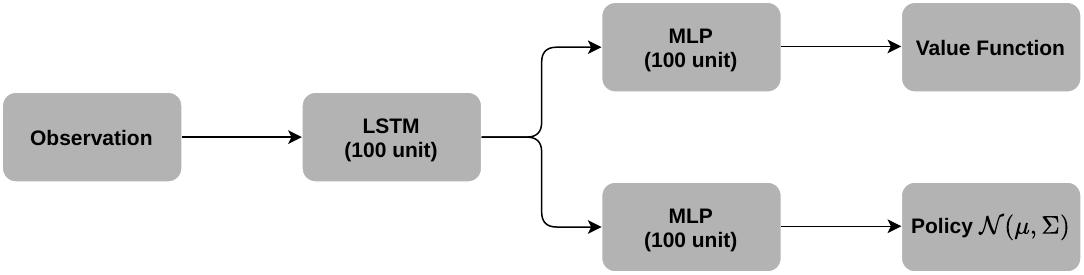}
	\caption{Network Architecture.}
	\label{fig: network architecture}
\end{figure}

Fig. \ref{fig: network architecture} shows the architectures of the neural network function approximator for the value function and policy.
Here, multivariate Gaussian distribution with diagonal covariance matrix is used as the stochastic policy.
With the estimated advantage function and the policy update rule in Eqn. \eqref{eq: ppo_update_rule}, the policy tends to converge to a suboptimal deterministic policy prematurely since a sequence of actions is required to change the intersection-crossing behavior due to its hierarchical nature.
Studies \cite{mnih2016asynchronous, williams1991function} show adding the entropy of the stochastic policy to the surrogate objective function effectively prevents such premature convergence.
As a result, the policy is updated by maximizing the following objective function:
\begin{gather}
L_{\text{mod},t}(\theta) = L_t(\theta) + \beta h(\mathcal{N(\mu,\Sigma)}). \label{eq: ppo_update_rule_with_entropy}
\end{gather}
Here, $L_t(\theta$) is the surrogate objective function defined in Eqn. \eqref{eq: ppo_update_rule}. $\beta$ and $h(\mathcal{N(\mu,\Sigma)})$ are the entropy coefficient and the entropy of the multivariate Gaussian policy.

Since PPO is on-policy, its sample efficiency is low. To accelerate the learning progress, multiple actors are distributed over different CPU processors for sample collection. Algorithm \ref{algorithm: ppo_for_deep_ecodriver} lists the detailed steps of the algorithm, and Appendix \ref{sec: hyperparameters} lists all the hyperparameters used for the training.

\begin{algorithm}[]
\SetAlgoLined
 Initialize EcoSim with the randomly generated routes\;
 Initialize the policy and value networks $\mu_\theta^{\pi_0}$, $V_\xi^{\pi_0}$\;
 \While{NOT converged}{
  \For{actor = $1, 2, ... N$}{
  Reset $v_{\text{vel},0} \sim \mathcal{U}(0,v_{\text{lim},0})$ and $SoC_0 \sim \mathcal{U}(0.3,0.8)$\;
  Execute policy $\pi_{\theta_{\text{old}}}$ in EcoSim for T time steps\;
  Record $(o_t, h_{o,t}, a_t, h_{a,t},r_t,o_{t+1}, h_{o,t+1})$ for each timestep\;
  Compute advantage estimates $\hat{A}^{\pi_{\theta_{\text{old}}}}(o_t,a_t)$ following Eqn. \eqref{eq: gae}\;
  }
  Compute the gradient of the surrogate objective function $L_\text{mod}(\theta)$ in Eqn. \eqref{eq: ppo_update_rule_with_entropy} and apply gradient ascent w.r.t $\theta$\;
  Update the value function and its parameter set $\xi$ following Eqn. \eqref{eq value_function_update}\;
  
 }
 \caption{PPO for Eco-driving}
 \label{algorithm: ppo_for_deep_ecodriver}
\end{algorithm}

\section{Methods for Benchmarking} \label{sec: methods for benchmarking}
To benchmark the fuel economy benefits of CAV technologies, it is crucial to establish a baseline representative of real-world driving.
In this work, the performance of the DRL agent is benchmarked against another two real-time control strategies and the wait-and-see solution.

\subsection{Baseline Controller} \label{sec: baseline}
The baseline controller consists of the Enhanced Driver Model (EDM), a deterministic reference velocity predictor that utilizes route features to generate velocity profiles representing varying driving styles \cite{gupta2019enhanced,gupta2020estimation}, and a rule-based HEV energy management controller.
The baseline strategy passes signalized intersections based on line-of-sight (LoS), a dynamic human-vision based distance parameter used to preview the upcoming route feature as devised by the Intersection Sight Distance (ISD) specified by the American Association of State Highway and Transportation Officials (AASHTO) and US DoT FHA \cite{aashto2001policy}.

\subsection{Online Optimal Controller} \label{sec: rollout}
In the previous work \cite{deshpande2021real}, a hierarchical MPC is formulated to co-optimize the velocity and powertrain controls with an aim to minimize energy in a 48V mild-HEV using Approximate Dynamic Programming (ADP).
The controller solves a long-term optimization at the beginning of the trip that evaluates a base policy using limited full-route information such as speed limits and positions of route markers such as stop signs, traffic lights.
To account for variability in route conditions and/or uncertainty in route information, a short-term optimization is solved periodically over a shorter horizon using the approximated terminal cost from the long-term optimization.
Time-varying SPaT information is accounted for by developing a heuristic controller that uses distance to the traffic light, current vehicle velocity to kinematically reshape the speed limit such that the vehicle can pass at a green light.
The formulation and details of the algorithm are referred to \cite{deshpande2021real}.
A detailed analysis is done to demonstrate the real-time implementation of the developed controller in \cite{deshpande2021vehicle} at a test-track in Columbus, OH.

\subsection{Deterministic Optimal Solution} \label{sec: DP_formulation}
The baseline controller, the online optimal controller, and the DRL agent can only preview the road information for the future 200m, which leaves the rest of the route information stochastic.

To show the sub-optimality from the control strategy and the stochastic nature of the problem, the wait-and-see solution, assuming the road information and SPaT sequence of the traffic lights over the entire routes are known \textit{a priori} are computed via Deterministic Dynamic Programming (DDP) \cite{sundstrom2009generic}.

To solve the problem via DDP, the state and action spaces are discretized, and the optimal cost-to-go matrix and the optimal control policy matrix are obtained from backward recursion \cite{olin2019reducing}.
Since each combination in the state-action space needs to be evaluated to get the optimal solution, the calculation of the wait-to-see solution for any trip used in the study can take hours with modern Central Processing Units (CPUs).
Here, the parallel DDP solver in \cite{zhu2021gpu} with CUDA programming is used to reduce the computation time from hours to seconds.  
The formulation and the details of the algorithm are referred to \cite{zhu2021gpu}.
Nevertheless, as the method requires the entire trip information \textit{a priori} and intense computation, the wait-and-see solution can only serve as a benchmark instead of a real-time decision-making module.

\section{Results} \label{sec: results} 
The training takes place on a node in Ohio Supercomputer Center (OSC) \cite{OhioSupercomputerCenter1987} with 40 Dual Intel Xeon 6148s processors and an NVIDIA Volta V100 GPU.
The results shown below requires running the training continuously for 24 hours. 

\begin{figure}[]
	\centering
	\includegraphics[width=\columnwidth]{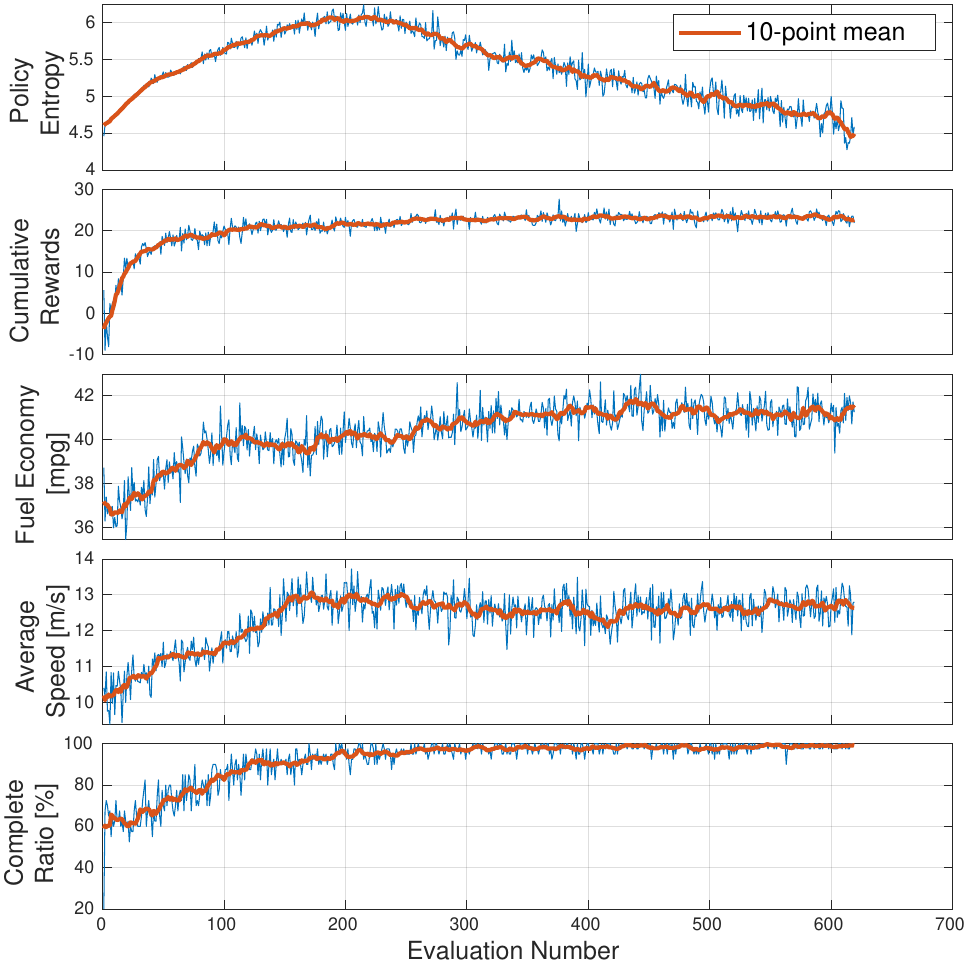}
	\caption{Evolution of Policy Entropy, Cumulative Rewards, Fuel Economy, Average Speed and Complete Ratio during Training}
	\label{fig: learning curve}
\end{figure}

As domain randomization is activated during training, the performance of the agent needs to be evaluated separately from the training to show the learning progress.
Here, 40 trips in the training set with domain randomization deactivated are executed for every 10 policy updates, i.e. every 4,000 rollout episodes.
In Fig. \ref{fig: learning curve}, the blue curves show the evolution of the mean policy entropy, the average cumulative rewards, the average fuel economy, the average speed and the complete ratio over the 40 randomly selected trips, and the red curves indicate the running average over 10 evaluation points.
Here, a trip is considered completed if the agent did not violate any constraint during the itinerary.
At the beginning of the training, the policy entropy increases to encourage exploration, and as the training evolves, the policy entropy eventually decreases as the entropy coefficient $\beta$ is linearly annealing.
On average, the agent reaches a performance with an average fuel economy of $41.0$ $mpg$ (miles per gallon) and an average speed of $\sim 12.7$ $m/s$.
The agent was able to learn to obey the traffic rules at signalized intersections within the operation design domain thanks to the properly designed rewarding mechanism. 
Although the use of negative rewards leads to the lack of safety guarantees, this does not pose a concern for an implementation since the handling of the long-tailed corner cases is usually deferred to a downstream controller module, see for instance \cite{sun2020optimal}. In addition, studies \cite{phan2022driving,vitelli2022safetynet} show that safety filters can be deployed on top of the superior planner to provide a safety guarantee.

\begin{figure}[!t]
    \centering
    \includegraphics[width=1\columnwidth]{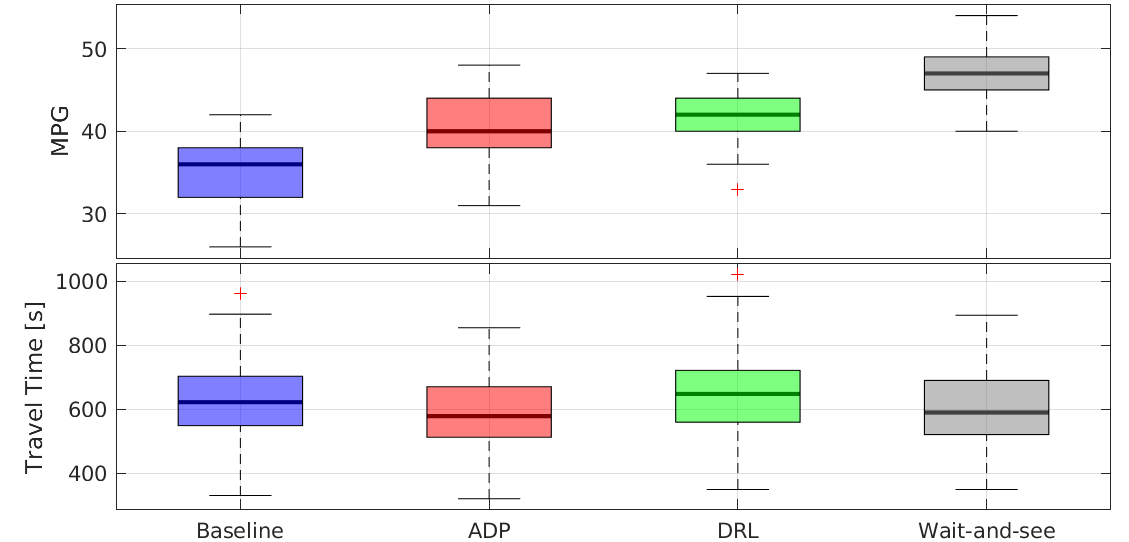}
    \caption{Fuel Economy, Travel Time Comparison and Charge Sustenance behavior for Baseline, DRL and Wait-and-see Solutions}
    \label{fig:MP_TT_Box_Plots}
\end{figure}

\begin{figure}[!t]
    \centering
    \includegraphics[width=1\columnwidth]{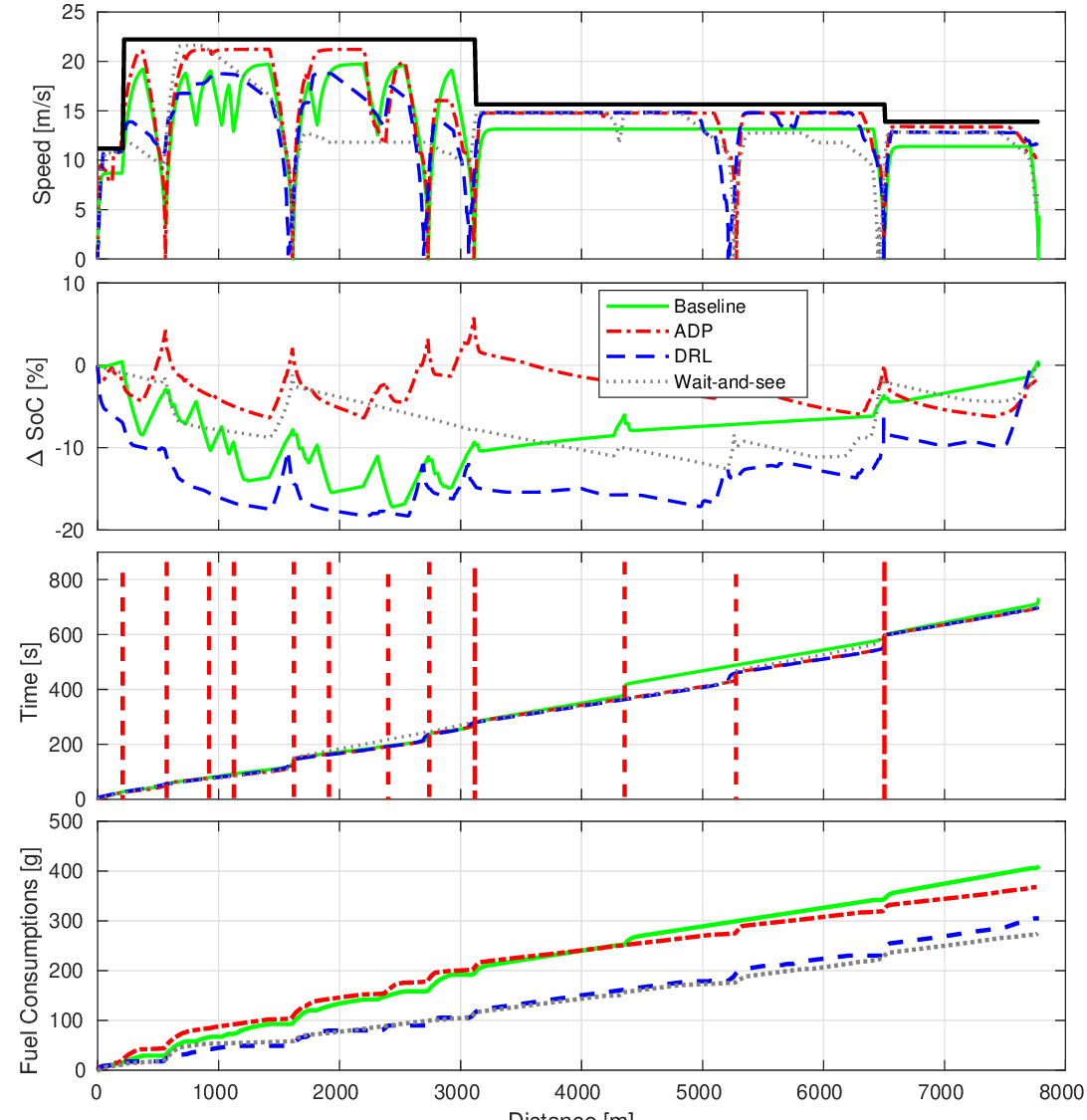}
    \caption{Comparison of Velocity, $\Delta SoC$, Time-Space and Fuel Consumption for Baseline, DRL and Wait-and-see Solutions [Urban Route]}
    \label{fig:Vel_SOC_Plots_urban}
\end{figure}

\begin{figure}[!t]
    \centering
    \includegraphics[width=1\columnwidth]{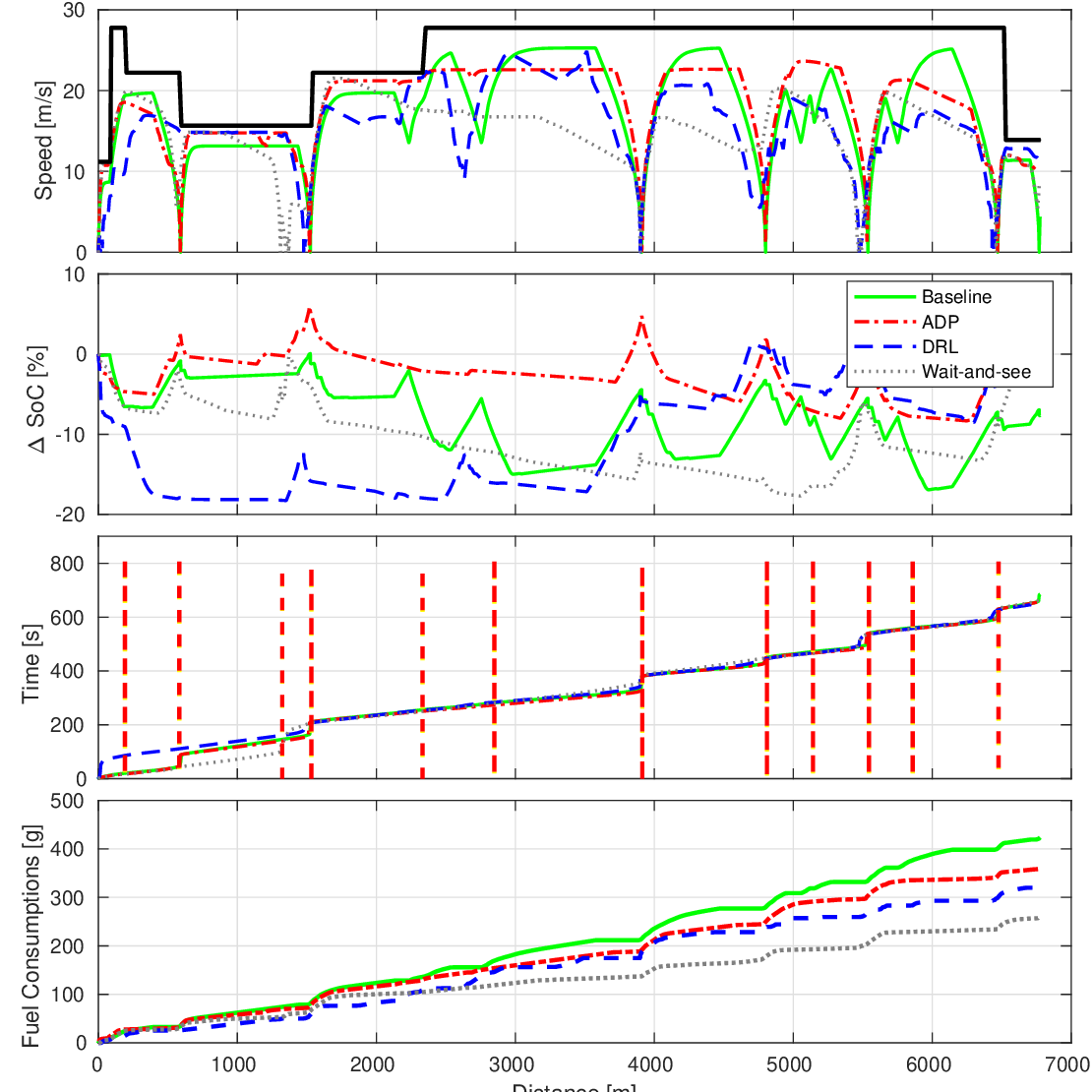}
    \caption{Comparison of Velocity, $\Delta SoC$, Time-Space and Fuel Consumption for Baseline, DRL and Wait-and-see Solutions [Mixed Route]}
    \label{fig:Vel_SOC_Plots_mixed}
\end{figure}

The performance of the DRL controller is then compared against the causal baseline and ADP controllers and the non-causal wait-and-see deterministic optimal solution among the 100 testing trips shown in Fig. \ref{fig:map_of_columbus}.
Fig. \ref{fig:MP_TT_Box_Plots} and Tab. \ref{tab: statistical_comparison} show the statistical comparison among the four strategies.
Here, the black line in each box represents the median of the data, and the lower and upper boundaries represents the $25^{\text{th}}$ and $75^{\text{th}}$ percentiles, respectively.
The extremes of the lines extending outside the box represent the minimum and maximum limit of the data and the “+” symbol represent the outliers.
With the comparable travel time, the DRL controller consumes $17.5\%$ less fuel over all trips compared to the baseline strategy. 
The benefits come from the more efficient use of the HEV powertrain by more energy recuperation into the battery during brake, the less use of the mechanical brake, and less unnecessary acceleration due to the presence of traffic lights.
These behaviors are demonstrated later in Fig. \ref{fig:Vel_SOC_Plots_urban} and \ref{fig:Vel_SOC_Plots_mixed}. 
Compared to the ADP controller, the DRL consumes $3.7\%$ less fuel, while being $\sim 1 m/s$ slower.
Meanwhile, considering the ADP controller requires solving the full-route optimization via DDP before departure and a trajectory optimization at every timestep in real-time, the DRL strategy, which requires only the forward evaluation of the policy network, is more computationally tractable.

\begin{figure}[!t]
    \centering
    \includegraphics[width=1\columnwidth]{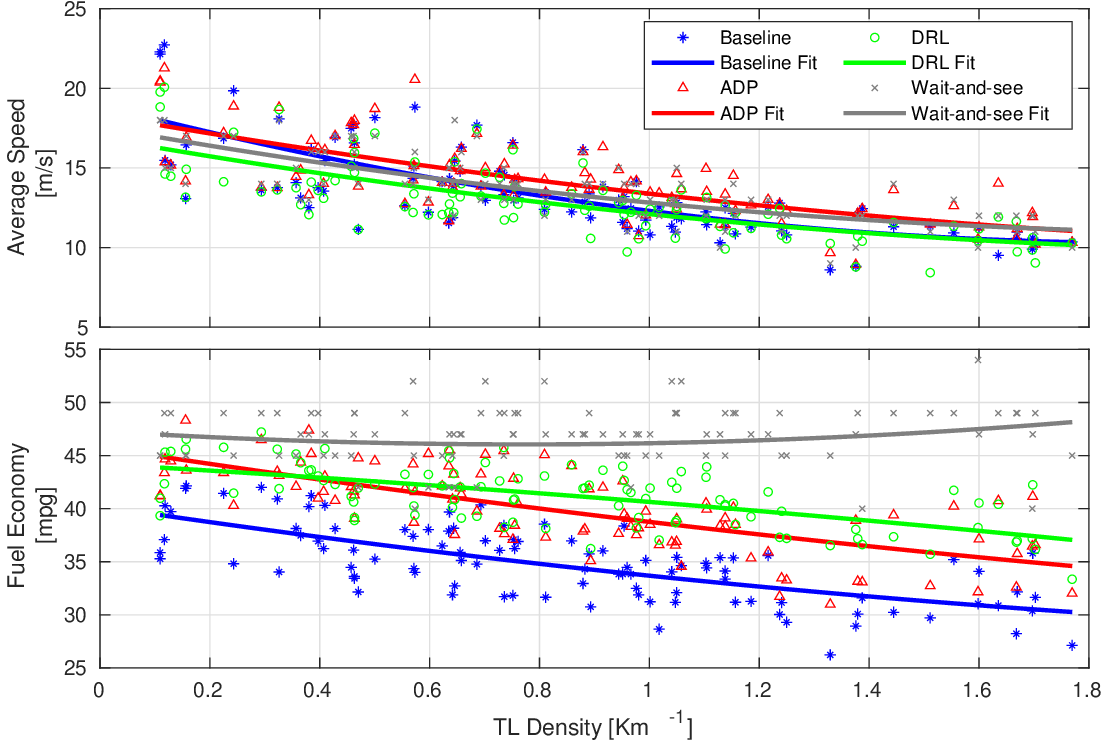}
    \caption{Variation of Average Speed and Fuel Economy against Traffic Light (TL) Density for Baseline, DRL and Wait-and-see Solutions}
    \label{fig:Vel_SOC_mean_Plots}
\end{figure}

The wait-and-see solution provides a dominant performance over the causal controllers. 
The additional benefits of the wait-and-see solution stem from the fact that the wait-and-see solution has the information of all the traffic lights over the entire trip, whereas the causal controllers only use the SPaT of the upcoming traffic light within 200m.
Such advantage is also reflected in Fig. \ref{fig:Vel_SOC_mean_Plots}.
Here, the average speed and the MPG of each trip are plotted against the traffic light density, i.e. the number of traffic lights divided by the total distance in kilometers.
Intuitively, the average speed of the three controllers decreases as the traffic light density increases.
Compared to the causal controllers, the fuel economy of the wait-and-see solution is not affected by the increase in traffic light density.
This is because the additional SPaT information from the subsequent traffic lights allows the wait-and-see solution to plan through the intersections more efficiently.
In the meantime, as suggested by the regression curves, the DRL controller is less sensitive to the increase in traffic light density than the baseline and ADP controllers.

\begin{table}[]
    \centering
    \caption{Fuel Economy, Average Speed and SoC Variance for Baseline, ADP, DRL and Wait-and-see Solutions}
    \label{tab: statistical_comparison}
    \begin{tabular}{c|c|c|c|c}
    \multicolumn{1}{l|}{} & \multicolumn{1}{c|}{Baseline} &\multicolumn{1}{c|}{ADP} & \multicolumn{1}{c|}{DRL} & \multicolumn{1}{c}{Wait-and-See} \\ \hline
    \begin{tabular}[c]{@{}c@{}}Fuel Economy\\ mpg\end{tabular} & 33.8 & 39.5 & 41.0 & 47.5\\ \hline
    \begin{tabular}[c]{@{}c@{}}Speed Mean\\ $m/s$\end{tabular} & 13.0 & 13.9 & 12.7 & 14.5\\ \hline
    \begin{tabular}[c]{@{}c@{}}$SoC$ Variance\\ $\%^2$\end{tabular}  & 13.2 & 21.6 & 18.2 & 22.6\\
    \end{tabular}
\end{table}

Fig. \ref{fig:Vel_SOC_Plots_urban} and \ref{fig:Vel_SOC_Plots_mixed}
show the trajectories of the three controllers in urban and mixed (urban and highway) driving conditions, respectively.
Driving under the same condition, the DRL controller was able to come to a full stop at signalized intersections less frequently compared to the baseline.
In addition, the DRL controller utilizes more of the battery capacity, i.e. a $SoC$ profile with higher variation, compared to the baseline.
DRL's efficient maneuvers while approaching the intersection coupled with better utilization of SoC results in up to 27\% reduction in fuel consumption for both urban and mixed driving when compared against the baseline.

\section{Conclusion} \label{sec: conclusion}
In this study, the eco-driving problem for HEVs with the capability of autonomously passing signalized intersections is formulated as a POMDP. To accommodate the complexity and high computational requirement of solving this problem, a learn-offline, execute-online strategy is proposed.
To facilitate the training, a simulation environment was created consisting of a mild HEV powertrain model and a large-scale microscopic traffic simulator developed in SUMO.
The DRL controller is trained via PPO with LSTM as the function approximators.
The performance of the DRL controller are benchmarked against a baseline strategy, a deterministic MPC strategy and a wait-and-see (optimal) solution.
With the properly designed rewarding mechanism, the agent learned to obey the constraints in the optimal control problem formulation. Furthermore, the learned explicit policy reduces the average fuel consumption by 17.5\% over 100 randomly generated trips in urban, mixed-urban and highway conditions when compared to the baseline strategy, while keeping the travel time comparable.

Future work will focus on three aspects. First, the hard constraint satisfaction will be rigorously analyzed. Then, the design of a reward function specific to the individual driver (personalization) will be investigated. Finally, the presence of a lead vehicle will be considered in the framework of an ecological ACC, which can be accomplished by expanding the state space and rewarding mechanism.

\appendices
\section{Reward Function Design}\label{sec: reward function design}
In general, the design of the reward function is iterative. To get the desired behavior from the trained policy, the numerical constants in the reward function typically requires tuning by humans.
Here, the numerical constants are listed Tab. \ref{tab:values_for_constants}. Some key takeaways are listed below.
\begin{enumerate}
    \item Normalize the scale of the reward function such that the numerical value is between $[-1, 1]$.
    \item The reward items $r_{\mathrm{vel}}$, $r_{\mathrm{batt}}$ and $r_{\mathrm{tfc}}$ are associated with the constraints, and they should be at least one order of magnitude higher than $r_{\mathrm{obj}}$.
    \item Rewards from the environment should reflect incremental incentives/penalties.
    For example, rewards associated with traffic lights are sparse, meaning that the agent receives these rewards periodically and needs a sequence of actions to avoid the large penalty.
    Eqn.\eqref{eq: traffic rewards} ensures the penalty for violating the traffic condition is proportional to how bad was the violation.
    \item Penalties related to constraints should be orders larger than those related to performance. 
\end{enumerate}

\begin{table}[]
\centering
\caption{Numerical Values of the Constants in Reward Function}
\label{tab:values_for_constants}
\begin{tabular}{c|c|c|c|c|c}
Constants & $c_{\text{obj}}$ & $c_{\text{time}}$ & $c_{\text{vel}}$ & $c_{\text{batt}}$ &  $c_{\text{tfc}}$\\ \hline
Values & $-0.001$ & $5$ & $\begin{bmatrix} -0.002 \\ -0.015 \end{bmatrix}$ & $\begin{bmatrix}
    -10 \\ -0.25
\end{bmatrix}$ & $\begin{bmatrix}
    -0.25\\-0.01\\-0.02
\end{bmatrix}$ 
\end{tabular}
\end{table}

\section{Hyperparameters}\label{sec: hyperparameters}
Two sets of hyperparameters have been obtained in this study. 
The first set contains those in the reward function, and the second contains those for the PPO algorithm.

The values in the first set are included in Tab. \ref{tab:values_for_constants}. In finding these values, the behavior of the agent was found to vary with $c_{\text{time}}$, as it governs the relative rewards between fuel consumption and travel time. For fair comparison against the other methods, the parameter was tuned to match the average speed while comparing fuel consumption. As for the reward coefficients related to the constraint penalties, the final performance does not appear sensitive as long as the penalties are large enough to dominate the positive rewards gained by ignoring the constraints while remaining numerically stable.

The hyperparameters used for training are listed in Tab. \ref{tab:hyperparameters}. The final performance was not found sensitive to this set of hyperparameters. The robustness here is most likely due to the fact that PPO is arguably one of the most stable model-free DRL algorithms, and that the parallel simulation environment provides a large amount of results efficiency.

\begin{table}[]
\centering
\caption{Numerical Values of the Constants in Reward Function}
\label{tab:hyperparameters}
\begin{tabular}{l|c}
Parameter & Value \\ \hline
LSTM Layer Size                 &  100      \\
MLP Layer Size                  &  100      \\ 
Samples per Episode             &  50       \\
Number of Episodes per Update   & 400       \\
Number of CPUs                  & 40        \\ 
Number of GPUs                  & 1         \\
PPO Clipping                    & 0.2       \\
GAE($\lambda$)                  & 0.8       \\
Truncated TD($\lambda$)          & 0.4       \\
Entropy Coefficient             & 0.01 (linear annealing)\\
Learning Rate $\alpha$          &  1e-6     \\ 
Discount Factor $\gamma$        & 0.999
\end{tabular}
\end{table}

\section*{Acknowledgment}
The authors acknowledge the support from the United States Department of Energy, Advanced Research Projects Agency–Energy (ARPA-E) NEXTCAR project (Award Number DE-AR0000794) and The Ohio Supercomputer Center.

\ifCLASSOPTIONcaptionsoff
  \newpage
\fi

\bibliographystyle{IEEEtran}
\bibliography{references}

\begin{IEEEbiography}[{\includegraphics[width=1in,height=1.25in,clip,keepaspectratio]{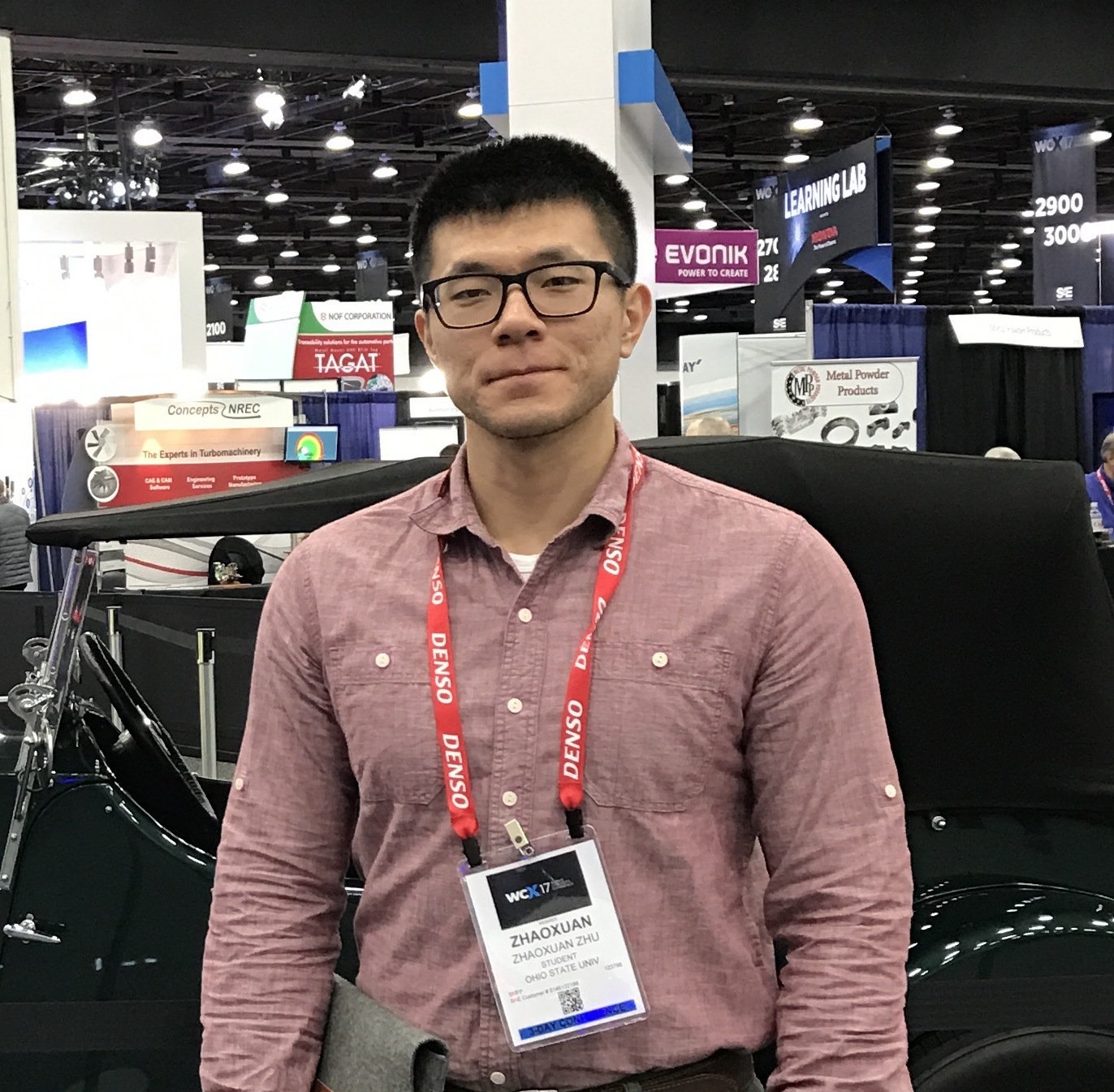}}]{Zhaoxuan Zhu}
received the B.Sc. degree (summa cum laude), M.S. degree and Ph.D. degree in Mechanical Engineering from The Ohio State University, Columbus, OH, USA in 2016, 2018, and 2021, respectively. He worked as a graudate research associate at the Center for Automotive Research from 2017 to 2021. He is currently a senior engineer at Motional. His research interests include optimal control, reinforcement learning, deep learning and their applications to connected and autonomous vehicles. 
\end{IEEEbiography}

\begin{IEEEbiography}[{\includegraphics[width=1in,height=1.25in,clip,keepaspectratio]{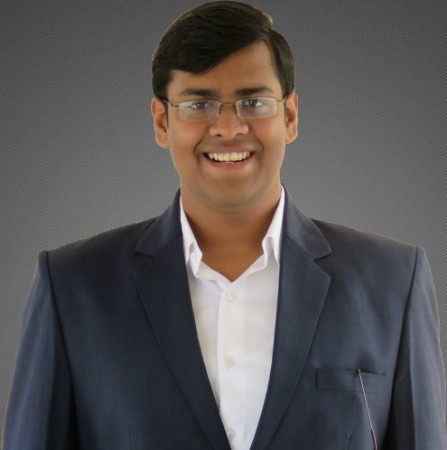}}]{Shobhit Gupta}
received the Bachelor of Technology degree in mechanical engineering from the Indian Institute of Technology Guwahati, Assam, India, in 2017, the M.S. and PhD degree in mechanical engineering from The Ohio State University, Columbus, OH, USA in 2019 and 2022 respectively. He is currently a Propulsion Controls Researcher at General Motors Global Research and Development, Detroit, MI, USA. His research interest include AI/ML and optimal control applicable to battery management systems, connected and autonomous vehicles and driver behavior recognition for predictive control.
\end{IEEEbiography}

\begin{IEEEbiography}[{\includegraphics[width=1in,height=1.25in,clip,keepaspectratio]{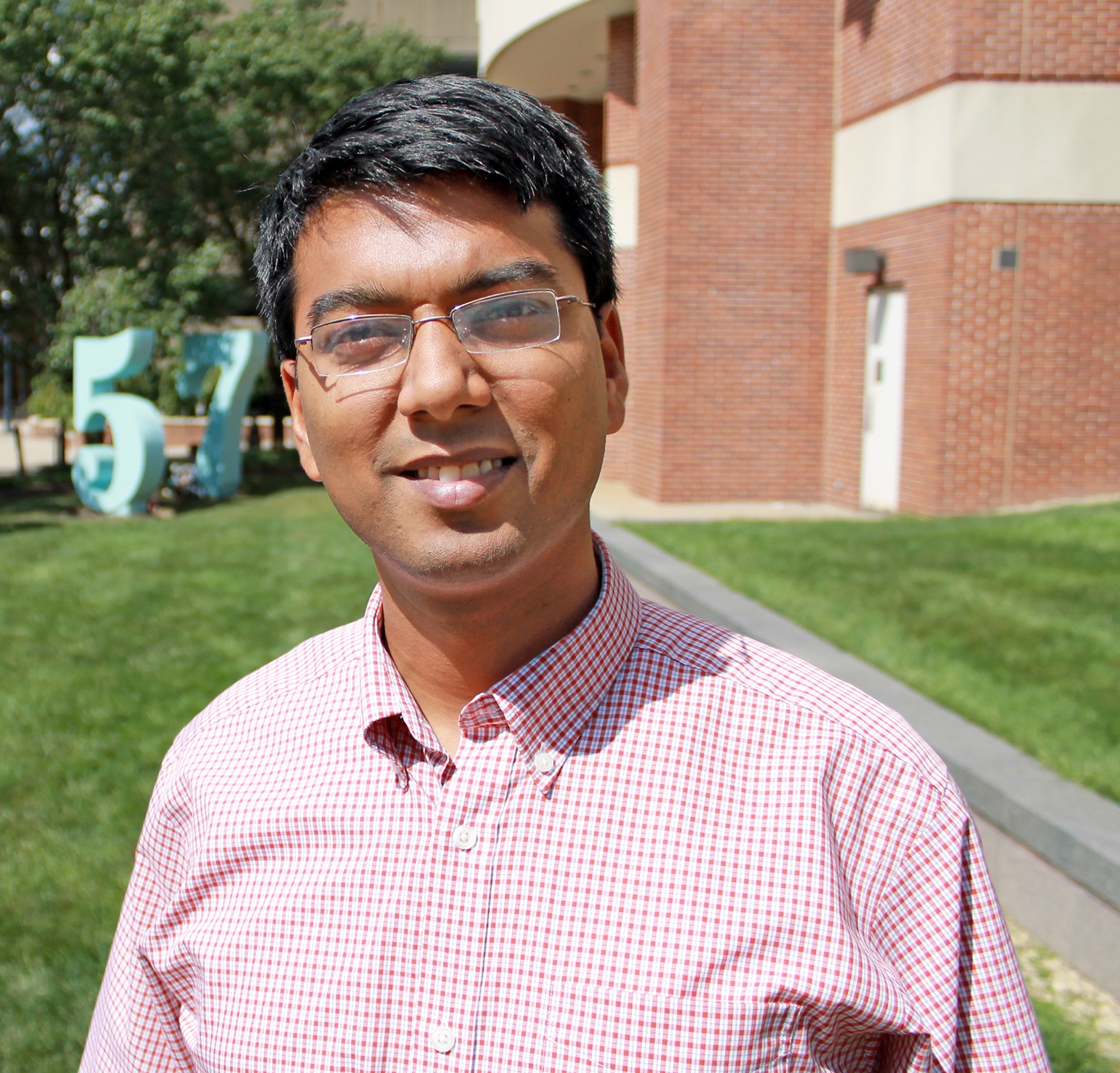}}]{Abhishek Gupta}
is an associate professor in the ECE department at The Ohio State University.
He completed his MS and PhD in Aerospace Engineering from University of Illinois at Urbana-Champaign (UIUC) in 2014, MS in Applied Mathematics from UIUC in 2012, and B.Tech. in Aerospace Engineering from IIT Bombay in 2009.
His research interests are in stochastic control theory, probability theory, and game theory with applications to transportation markets, electricity markets, and cybersecurity of control systems.
\end{IEEEbiography}

\begin{IEEEbiography}[{\includegraphics[width=1in,height=1.25in,clip,keepaspectratio]{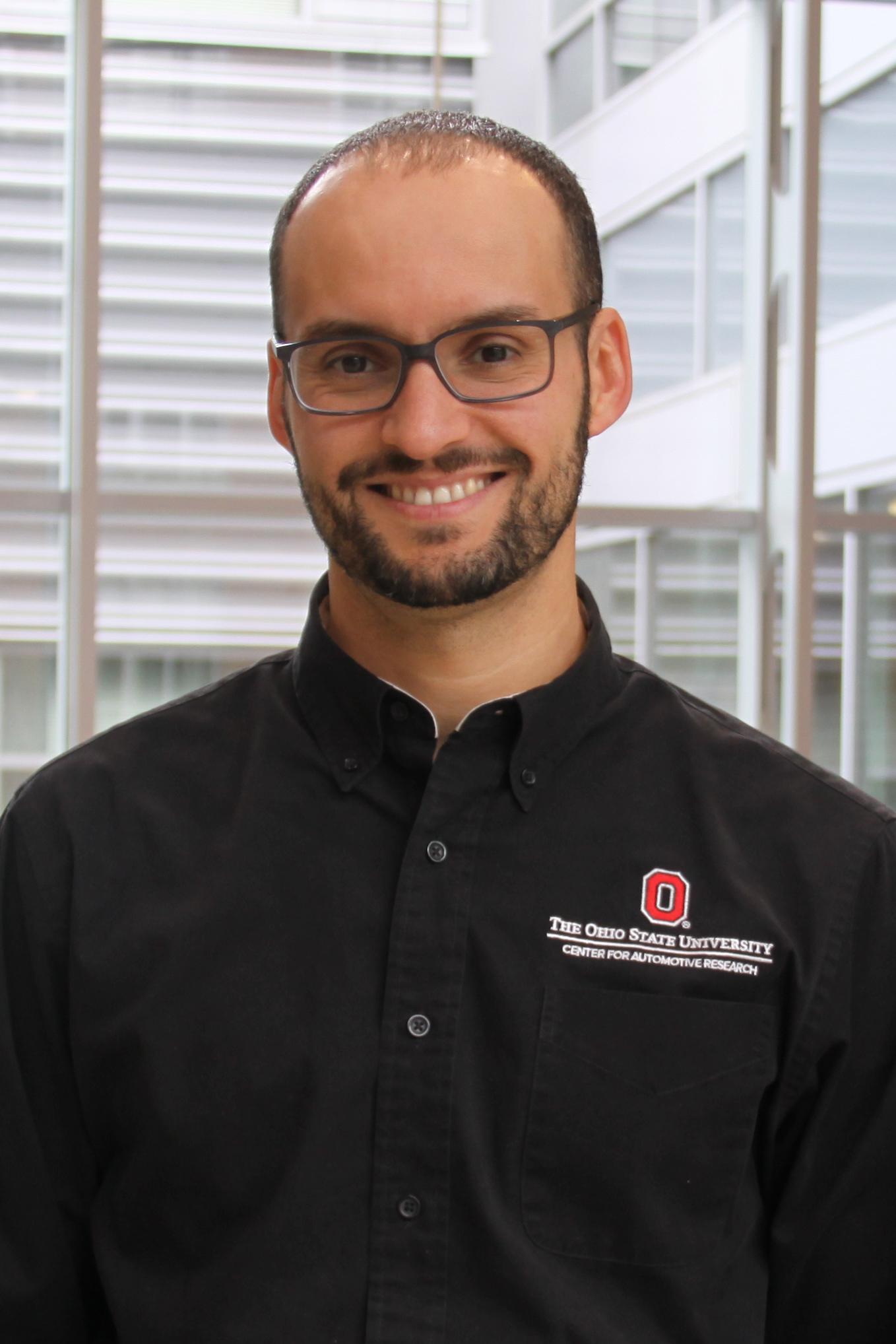}}]{Marcello Canova}
received the Ph.D. degree in mechanical engineering from the University of Parma, Parma, Italy. He is currently a Professor of mechanical and aerospace engineering with The Ohio State University, Columbus, OH, USA and the Associate Director of the Center for Automotive Research.

His research focuses on energy optimization and management of ground vehicle propulsion systems, including internal combustion engines, hybrid-electric drivetrains, energy storage systems, and thermal management. He is the author of more than 150 journal and conference articles, five U.S. Patents and he received, among others, the SAE Vincent Bendix Automotive Electronics Engineering Award in 2009, the SAE Ralph E. Teetor Educational Award in 2016, the NSF CAREER Award in 2016, the Lumley Research Award in 2012, 2016, and 2020, and the Michael J. Moran Award for Excellence in Teaching in 2017.
\end{IEEEbiography}

\end{document}